\documentclass[fleqn,usenatbib]{mnras}

\usepackage{newtxtext,newtxmath}

\usepackage[T1]{fontenc}

\DeclareRobustCommand{\VAN}[3]{#2}
\let\VANthebibliography\thebibliography
\def\thebibliography{\DeclareRobustCommand{\VAN}[3]{##3}\VANthebibliography}

\usepackage{graphicx}	
\usepackage{amsmath}	
\usepackage{xcolor}

\title{A comparison of G-band brightness as a proxy-magnetometer in various magnetic configurations}

\author[Shukla et al.]{
Malay Shukla,$^{1, 2}$
Sneha Pandit,$^{3}$
Nitin Yadav,$^{1,2,4}$\thanks{E-mail: nitnyadv@gmail.com}
\\
$^{1}$School of Physics, Indian Institute of Science Education and Research Thiruvananthapuram, Thiruvananthapuram 695551, Kerala, India\\
$^{2}$Center for High-Performance Computing, Indian Institute of Science Education and Research Thiruvananthapuram, Thiruvananthapuram 695551,
Kerala, India\\
$^{3}$Inter-University Centre for Astronomy and Astrophysics, Post Bag 4, Ganeshkhind, Pune 411007, Maharashtra, India\\
$^{4}$Department of Physics, Indian Institute of Technology Delhi, Delhi 110016,
 India
}

\date{Accepted XXX. Received YYY; in original form ZZZ}

\pubyear{\the\year{2025}}

\begin{document}
\label{firstpage}
\pagerange{\pageref{firstpage}--\pageref{lastpage}}
\maketitle

\begin{abstract}
We investigate the diagnostic potential of the G-band at 430.4\,nm for probing small-scale magnetic fields in the solar photosphere. Combining three-dimensional MHD simulations from the MURaM code and spectral synthesis via the RH 1.5D code, we evaluate the intensity contrast in the G-band filtergrams by comparing the filter centered at 430.4\,nm in comparison to the conventional 430.5\,nm. Our results show that filtergrams centered at 430.4\,nm provide higher contrast across varying magnetic environments, particularly at narrow filter widths. This enhancement arises from its slightly higher formation height and greater sensitivity to temperature variations in magnetized regions. These findings indicate that G-band filtergrams centered at 430.4 nm show enhanced diagnostic potential under the assumptions of the present modeling. The obtained results are also relevant and suggest potential applications in stellar contexts, where molecular bands are often used as proxies for magnetic activity.

\end{abstract}




\section{Introduction}
\label{sec:introduction}
The solar photosphere hosts a range of magnetic structures, from large sunspots to small-scale magnetic flux tubes \citep{Solanki_1993, Solanki_2006, Hood_2011}.
Even in apparently quiet regions, high-resolution observations have revealed a complex network of magnetic elements evolving continuously in intergranular lanes \citep{de_Wijn_2008, Lites_2009}. 
Understanding solar magnetism is essential because large-scale fields drive major eruptive events, such as flares and coronal mass ejections, while numerous small-scale magnetic heating events have been shown by recent high-resolution observations to play an equally, if not more, significant role in powering the outer solar atmosphere.
This recognition has sparked growing interest in studying small-scale magnetic features in greater detail, as they contribute directly to the magnetic energy budget and to the heating of the chromosphere and corona through processes such as magnetic reconnection and magnetohydrodynamic (MHD) wave propagation \citep{schrijver1998, Gošić_2014}.


The detection and characterisation of small-scale magnetic fields rely on a combination of spectropolarimetric and imaging diagnostics.
Fe I lines in the visible spectrum, particularly around 630 nm, have been extensively used for spectropolarimetric studies of photospheric fields with instruments such as SDO/HMI, SST/CRISP, and more recently, DKIST \citep{Scharmer_2008, Scherrer_2012, Rimmele_2020}.
Other diagnostics, such as Si I lines in the near-infrared, also provide valuable information on magnetic field strengths and stratification in the photosphere \citep{Centeno_2009, Shchukina_2017}. Alongside these atomic lines, the G-band has been widely employed as a proxy for small-scale magnetic concentrations because of its ability to produce high-contrast images of magnetic bright points compared to the nearby continuum \citep{Berger_Title_1995, Sanchez_2004}.
Several studies have quantitatively compared the contrast of G-band bright points with that observed in Fe I continuum or other spectral diagnostics, consistently demonstrating the enhanced visibility of magnetic elements in the molecular band \citep{Utz_2013, Jess_2012}.





Early theoretical work by \citep{Uitenbroek_2004} suggested that centering the G-band filter at 430.4 nm, rather than the conventional 430.5 nm, could provide enhanced sensitivity to magnetic fields due to the contribution of specific CH molecular transitions.
However, this proposition was not explored further in detail, primarily because observational strategies and instrumental designs continued to rely on the traditional 430.5 nm setting.
Moreover, the use of the 430.4 nm wavelength required narrowband filters with high spectral resolution, which were technically challenging to implement at that time.


With the advent of realistic three-dimensional MHD simulations and advanced spectral synthesis techniques, it is now possible to re-examine the diagnostic potential of the 430.4 nm band.
The motivation of the present study is to assess whether shifting the central wavelength of G-band filters to 430.4 nm, as originally suggested by \citet{Uitenbroek_2004}, offers a tangible improvement for the detection of small-scale magnetic structures in the solar photosphere.

The G-band is densely populated with numerous absorption lines formed by CH molecules, which play a crucial role in shaping the observed brightness features in this spectral region.
Understanding brightness enhancements in the G-band requires examining the physical conditions of the solar photosphere and their influence on CH molecule abundance. CH molecules produce strong absorption features in the photosphere, and variations in G-band brightness are primarily governed by the opacity of these molecular lines \citep{steiner_bruls_2001, Schussler_2003}.
Studies have shown that in regions of increased magnetic field strength, CH molecules dissociate more efficiently, reducing the local molecular opacity and thereby enhancing the continuum intensity \citep{Shelyag_2004}.

Convective motions at the solar surface transport magnetic fields into cooler intergranular lanes, where plasma flows downward.
In these regions, magnetic fields become concentrated into flux tubes in which magnetic pressure exceeds gas pressure. 
To maintain pressure balance with the surroundings, the internal gas pressure within these flux tubes is reduced, leading to partial evacuation of plasma in these regions and the formation of a Wilson depression, wherein the $\tau = 1$ surface is shifted to deeper, hotter layers \citep{Spruit_1976, steiner_bruls_2001}.
This evacuation of plasma lowers the CH number density, further decreasing opacity and allowing radiation from beneath the normal photospheric layer to emerge.
Additional heating can result from radiative transfer from the surrounding hotter granules into the sides of evacuated flux tubes, as found in numerical models of solar magneto-convection \citep{Carlsson_2004}.
As a result, magnetic elements appear brighter in the G-band, a phenomenon confirmed by both high-resolution observations and radiative MHD simulations \citep{steiner_bruls_2001, Schussler_2003, Schussler_Vogler_2003, Shelyag_2004, Beck_2007, criscouli_uitenbroek_2014}.

In addition to reduced gas density, elevated temperatures within magnetic flux tubes also play a critical role in enhancing brightness.
The suppression of convective energy transport in these regions allows radiative heating from the surrounding "hot walls" to increase the local temperature, as energy is transported laterally into the partially evacuated flux tubes \citep{Spruit_1976, Carlsson_2004}.
These higher temperatures promote the further dissociation of CH molecules, which have a relatively low dissociation energy of 3.47 eV, making them highly susceptible to destruction by thermal and radiative effects in magnetised regions \citep{Knoelker_1991, Almeida_2001, Uitenbroek_Trits_2006}. 

The absorption features observed in the G-band primarily originate from CH molecular transitions, particularly electronic transitions between the $A^2\Delta$ and $X^2\Pi$ states.
These transitions, along with their rotational-vibrational structure, are highly sensitive to local thermodynamic conditions and strongly affect the depth and shape of absorption features \citep{Jorgensen_1996, berdyugina_solanki_2002, Berdyugina_2003}.
Consequently, variations in CH molecule abundance, driven by temperature and density fluctuations, directly impact the observed intensity.
Because of these physical mechanisms, intensity-based imaging has proven to be particularly effective.
In particular, G-band bright points, localised brightness enhancements in the 430-431 nm region dominated by CH molecular absorption, have been widely used as proxies for small-scale magnetic flux concentrations \citep{Muller_Roudier_1984, Keller_1992, Berger_1998, Berger_Title_1995, Utz_2009}

Several observational and modeling efforts have confirmed the utility of the G-band in magnetic field diagnostics \citep{Berger_Title_1995, Keller_1992, Berger_1998, Schussler_2003, Shelyag_2004, Utz_2009}. In particular, \citet{Uitenbroek_2004} suggested that shifting the filter center from the traditional 430.5 nm to 430.4 nm may provide enhanced sensitivity.

In this study, we investigate the intensity contrast in the G-band using synthetic observations generated from realistic MHD simulations.
By comparing the brightness response at 430.4 nm and 430.5 nm in various magnetic configurations and filter widths, we assess the suitability of 430.4 nm as a better diagnostic wavelength for detecting small-scale magnetic structures in the solar photosphere.


The paper is organised into the following sections. In Section ~\ref{sec:methods}, we describe the numerical simulations and spectral synthesis tools used in this study, including details of the MURaM radiative MHD code and the RH 1.5D radiative transfer calculations.
Section ~\ref{sec:results} presents the main results, focussing on the comparative analysis of intensity contrast at 430.4 nm and 430.5 nm across different magnetic field regions and spectral resolutions.
In Section ~\ref{sec:discussions}, we discuss the physical mechanisms underlying the observed contrast differences and their implications for solar magnetic diagnostics.

\section{Methods}
\label{sec:methods}
\subsection*{MHD Simulations}
To simulate the realistic solar conditions for Quiet Sun (QS), Weak Plage (WP) and Strong Plage (SP), we used the MURaM (MPS/University of Chicago Radiative MHD) code \citep{Vogler_2005} i.e. a well-established three-dimensional radiative magnetohydrodynamics code widely used to study various solar atmospheric phenomena \citep{Rempel_2016, Yadav_2020, Przybylski_2022, kannan_2024}. 
MURaM code incorporates non-grey radiative transfer and a realistic equation of state, enabling self-consistent coupling between convection and magnetic fields. 
It has been extensively used in studies of magneto-convection and the evolution of small-scale solar magnetic fields \citep{Rempel_2014}, as well as flux emergence processes \citep{Cheung_2008}.


Our simulation domain spans $12\,\mathrm{Mm} \times 12\,\mathrm{Mm} \times 4\,\mathrm{Mm}$, extending from $-1.5$\,Mm below the photosphere to $+2.5$\,Mm above it in the vertical ($z$) direction.
The grid spacing is $10$\,km per cell in all directions, allowing us to resolve fine structures such as intergranular lanes and magnetic bright points.
Periodic side boundaries and an open bottom boundary allow mass-conserving inflows and outflows, which are essential for capturing realistic convective behaviour.

We simulated three distinct magnetic regions of the solar photosphere.
All simulations were initialised as purely hydrodynamical runs without magnetic fields and were evolved until they reached a convective quasi-steady state.
Subsequently, a uniform vertical magnetic field was introduced, and each run was further evolved until a statistically stationary state was attained \citep{Vogler_2005, Rempel_2014}.
The first setup corresponds to a QS environment with weak internetwork fields and an average vertical magnetic field of 10 Gauss (G), consistent with spectropolarimetric observations that report predominantly weak fields in internetwork regions \citep{Orozco_Su_rez_2007, Lites_2008, Danilovic_2016}. 
The second setup represents a WP region with an average vertical field of 50 G, in agreement with observations showing that plage regions contain roughly an order-of-magnitude larger vertical flux than the QS \citep{Ishikawa_2009}. 
The third setup represents a SP region with an average vertical field of 200 G, capturing the presence of kilo-Gauss magnetic flux concentrations, as reported in the literature \citep{Buehler_2015, Buehler_2019, Kahil_2019, Liu_2025}.
From each simulation, representative snapshots were selected for analysis, focusing on the impact of magnetic fields on CH molecule dissociation and the resulting intensity contrast in the G-band. 
These simulation snapshots are then further explored to investigate the connection between small-scale magnetism and the radiative signatures observed in high-resolution G-band filtergrams.


\begin{figure*}
    \centering
    
    \includegraphics[width=0.95\linewidth]{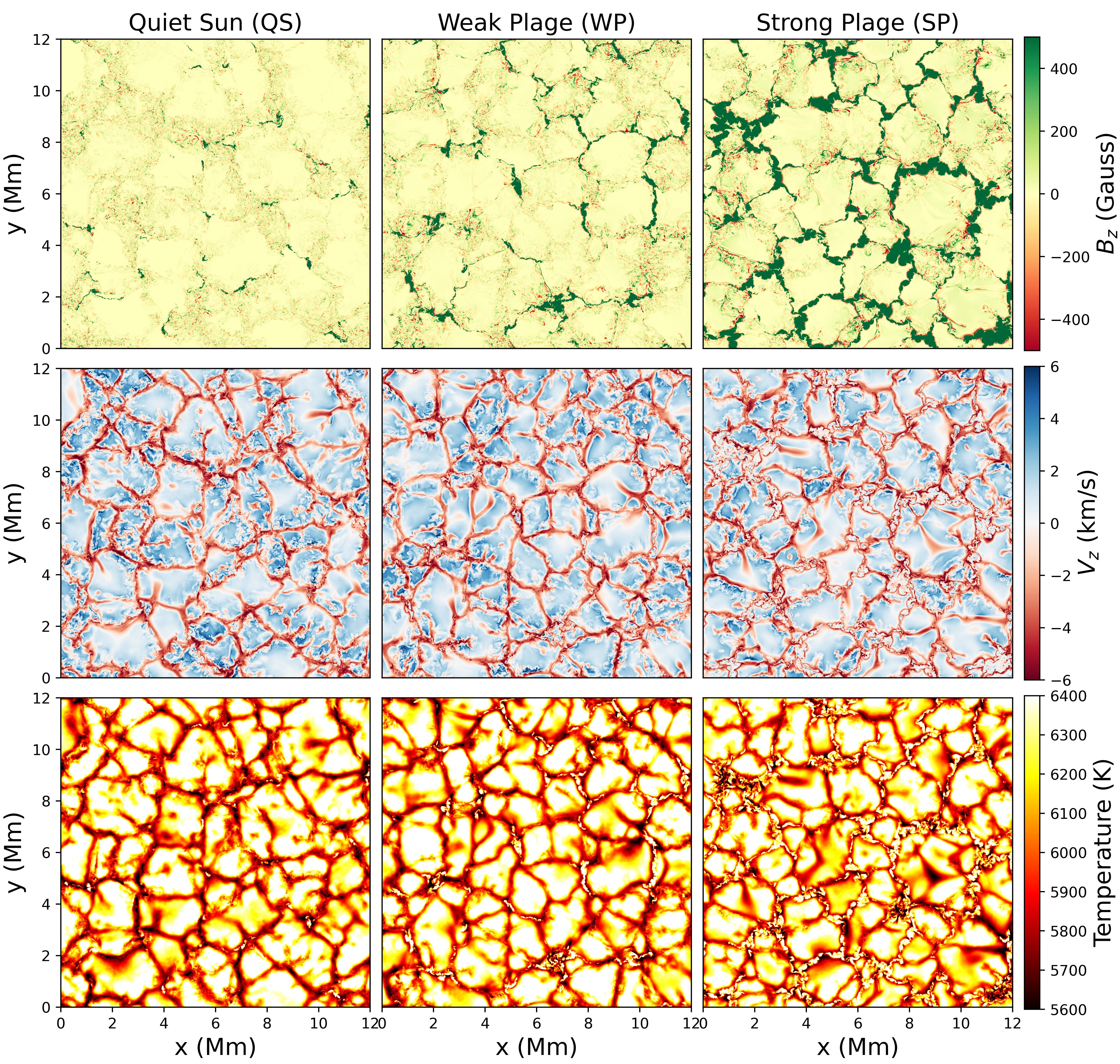}
    
    \caption{Maps of magnetic field strength ($B_z$), line-of-sight velocity ($v_z$), and temperature (T) at the $\tau = 1$ layer for three simulation setups corresponding to QS, WP, and SP regions.. 
}
    \label{fig:physical_quantities}
\end{figure*}

\begin{figure*}
    \centering
    \includegraphics[width=0.9\linewidth]{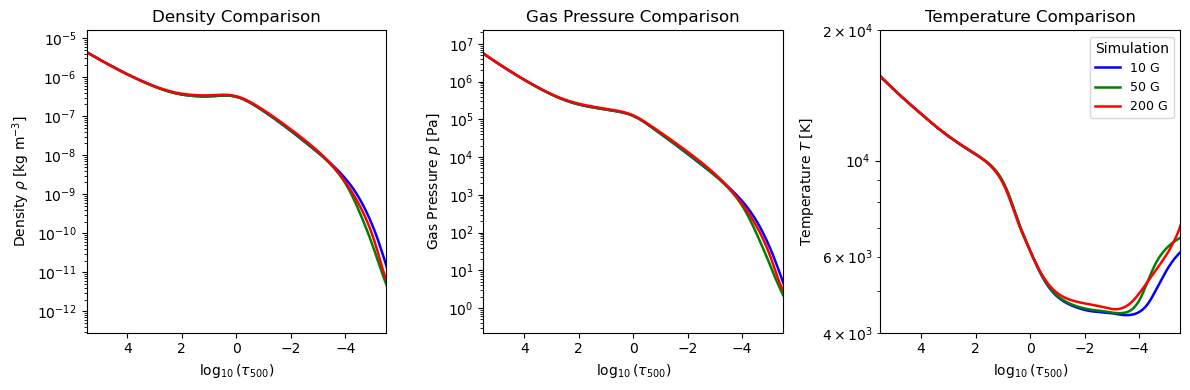}
    \caption{Spatially averaged profiles of density ($\rho$), gas pressure ($p$), and temperature ($T$) as functions of log($\tau_{500}$) for the three simulation setups. The averaging is performed over the full horizontal extent (12 Mm $\times$ 12 Mm) of each domain. The plotted range in log($\tau_{500}$) corresponds to the atmospheric layers between approximately -0.5 Mm to +1.0 Mm in geometric height, the region used as input for the RH 1.5D radiative transfer calculations. }
    \label{fig:average_quantities}
\end{figure*}

\subsection*{Spectral Synthesis}
For the spectral synthesis calculations, we used the RH 1.5D radiative transfer code \citep{Uitenbroek_RH_2001, Tiago_Periera_RH_2015}, a widely used tool for generating synthetic spectra from realistic atmospheric models. 
RH 1.5D solves the radiative transfer equation under both local thermodynamic equilibrium (LTE) and non-LTE conditions, making it suitable for spectral analysis in the solar photosphere. 
The code accounts for detailed atomic and molecular line formation physics and includes opacity contributions from multiple species, which are essential for accurately synthesizing the CH-dominated G-band.

In the present study, we synthesized spectra over the 430–431 nm wavelength range, sampling 1000 points with uniform spacing. 
This interval encompasses a dense forest of CH molecular transitions and corresponds to a well-characterized region in high-resolution solar observations \citep{BASS2000}. 
For the spectral synthesis, we included CH molecular transitions together with hydrogen (H) and molecular hydrogen (H$_2$).
The input atmosphere consists of three-dimensional snapshots from the MURaM simulations, containing thermodynamic variables, which are treated as independent vertical columns.

To simulate realistic observational conditions, the synthesized spectra were convolved with Gaussian filters representing different spectral resolutions. Specifically, we applied Full-Width Half Maximum (FWHM) values of 0.5 nm, typical of broadband instruments such as DKIST/VBI and 0.01 nm to mimic high-resolution narrowband filters.
Previous work by \citep{Uitenbroek_2004} proposed that a filter width of about 0.02 nm would be required to exploit the diagnostic potential of the 430.4 nm region for magnetic measurements.
It should be noted that in this study we consider only spectral degradation through Gaussian convolution and do not include spatial degradation due to the telescope point spread function (PSF), as our focus is on isolating the effects of spectral resolution on the magnetic diagnostics.


The resulting synthetic spectra provide key insights into how magnetic field concentrations modulate brightness in the G-band. By combining high-resolution radiative MHD simulations with detailed molecular radiative transfer, our analysis improves the understanding of CH molecule dissociation and its impact on intensity contrast in magnetized regions of the solar photosphere.

\section{Results}
\label{sec:results}

\begin{figure*}
    \centering
    \includegraphics[width=\linewidth]{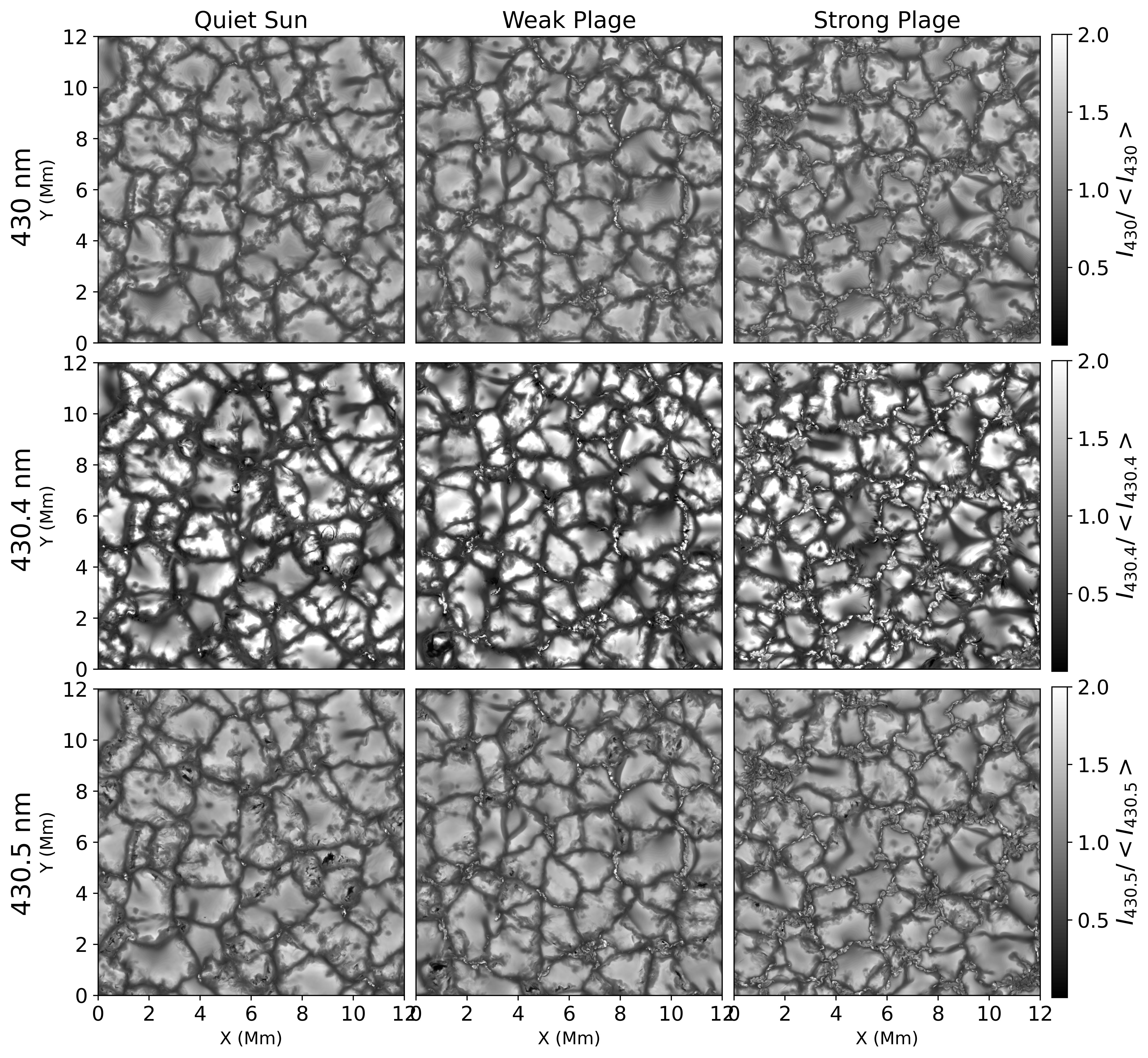}
    \caption{Brightness intensity maps for three different simulation setups (columns). The top row shows the continuum intensity at 430 nm, emphasising the overall photospheric brightness distribution. The middle row displays the intensity map at 430.4 nm, where the intergranular lanes and granules exhibit higher contrast. In comparison, the bottom row illustrates the intensity map at 430.5 nm, showing different intensity distributions between intergranular lanes and granules.}
    \label{fig:different intensities}
\end{figure*}


Figure~\ref{fig:physical_quantities} illustrates key physical quantities derived from the MHD simulations.
These outputs provide essential insight into the thermal and magnetic structure of the solar photosphere and serve as input for the subsequent spectral synthesis.

Fig. \ref{fig:physical_quantities} shows the vertical component of the magnetic field ($B_z$), the line-of-sight velocity ($v_z$), and the temperature (T) at the $\tau_{500}=1$ surface for the three simulations.
The QS case displays weak, intermittent magnetic field concentrations confined to intergranular lanes. 
In contrast, the WP simulation exhibits more organized distributions of kilogauss-strength flux tubes, while the SP features highly magnetized structures with field strengths reaching up to $\sim 3300 G$.

This accumulation of field in the first row of the Fig~\ref{fig:physical_quantities} results from plasma compression in the lanes, where converging flows concentrate magnetic flux.
Such flux intensification is well established in magneto-convection, where the combined action of convective downflows and magnetic tension leads to stable flux tube formation \citep{Stein_2006, Rempel_2014}. 
Evacuation of plasma within these magnetic elements lowers internal gas pressure, enhancing field strength and resulting in nearly evacuated tubes with high magnetic energy density, consistent with theoretical models \citep{Spruit_1976, Deinzer_1984a, Deinzer_1984b}.

The second row of Fig.~\ref{fig:physical_quantities} shows the vertical velocity field ($v_z$) at $\tau_{500} = 1$, highlighting the convective flow pattern in each setup. Bright granule interiors are dominated by upflows, while intergranular lanes host strong downflows.
These downflows are essential for the intensification of magnetic flux tubes, particularly in the WP and SP cases, where velocities reach several km/s.
The behavior results from the interaction between convective downflows surrounding the magnetic structures, which concentrate magnetic flux, and the suppression of vertical convective motions by magnetic pressure, leading to intensified flux tubes \citep{Deinzer_1984a}.
These thermodynamic and magnetic effects are discussed in detail in section \ref{sec:introduction}. This sequence of effects is a central feature of flux tube models and has been extensively studied in radiative heating analyses \citep{Spruit_1976, Knoelker_1991}.

The third row of Fig.~\ref{fig:physical_quantities} shows the temperature distribution at $\tau_{500} = 1$. As expected, temperatures are highest in granular upflows and lowest in the intergranular lanes. In the intergranular lanes, small-scale regions of elevated temperature are observed that coincide with strong magnetic field concentrations seen in the first row, illustrating the hot wall effect, where the sidewalls of the flux tubes are heated by radiation from surrounding granules. In the WP and SP snapshot, some locations show cooler plasma within the central parts of the magnetic concentrations, which is more prominent in the SP regions. These temperature patterns arise primarily because high magnetic pressure suppresses convective energy transport, inhibiting overturning motions and causing convective stagnation. In larger flux concentrations, the hot wall effect becomes less efficient because the increased tube width limits the lateral penetration of granule radiation \citep{Riethmuller_2017}. These flux tubes act as a conduit for the transport of radiation energy.

Figure~\ref{fig:average_quantities} presents the domain-averaged profiles of density ($\rho$), gas pressure ($p$), and temperature (T) as functions of height.
These stratifications provide complementary context to the maps in panel (a), highlighting how the underlying thermodynamic structure differs between the QS, WP, and SP cases and influences the subsequent spectral synthesis.




This explains why the plage simulations, which contain strong, well-formed magnetic flux tubes, exhibit a more pronounced temperature enhancement compared to the QS case. The combined effects of reduced molecular opacity, suppressed convective energy transport, and lateral radiative heating fundamentally shape the thermal structure of magnetized regions. The resulting brightness enhancements observed in the G-band are a direct manifestation of these processes, underscoring the band’s diagnostic value for identifying and analyzing small-scale magnetic field concentrations in the solar photosphere.



Fig.~\ref{fig:different intensities} shows the spatial distribution of intensity at the three selected wavelengths for the three simulation setups. 
The different rows correspond to 430 nm continuum, 430.4 nm, and 430.5 nm respectively, while the different columns represent QS, WP, and SP regions.
The 430 nm map corresponds to a pure continuum wavelength, chosen without any CH absorption lines, and serves as a reference for the line-core maps shown at 430.4 nm and 430.5 nm.
It is evident from the figure that the intensity contrast is consistently highest at 430.4 nm compared to both the nearby continuum and the more commonly used G-band (filter centered at 430.5 nm).
These results suggest that both the central wavelength and spectral bandwidth of the imaging filter significantly influence diagnostic sensitivity \citep{Uitenbroek_Trits_2006}.
In contrast, 430.4 nm retains greater detail and sensitivity to fine-scale variations, emphasizing its diagnostic potential for small scale magnetic structures \cite{Uitenbroek_2004}.

The enhanced contrast observed at 430.4\,nm indicates that this wavelength is particularly well-suited for detecting magnetic bright points and for probing the thermodynamic conditions in magnetized regions. The improved sensitivity arises from the interplay of CH molecular transitions, which respond strongly to changes in opacity, temperature, and gas density in the presence of magnetic fields.


\begin{figure*}
    \centering
    \includegraphics[width=\linewidth]{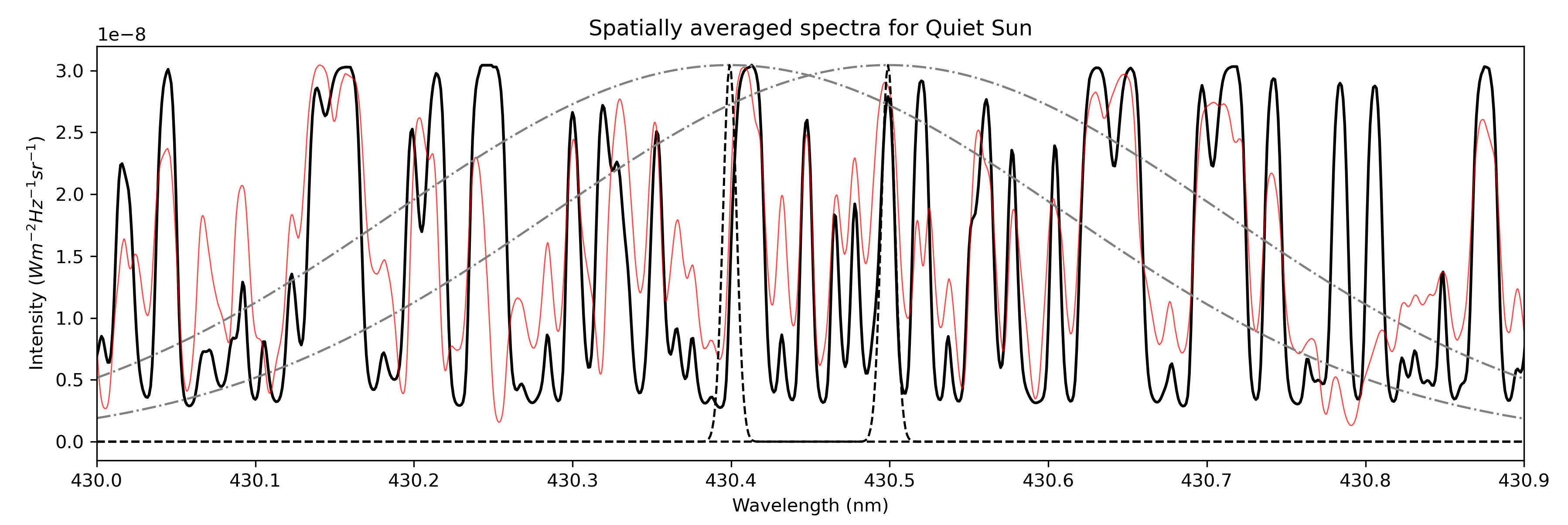}
    \caption{A plot showing the spatially averaged G-band spectrum (solid black line) for the QS simulation, compared with the reference spectrum from the ATLAS database presented with red line. Overplotted are Gaussian filter profiles representing different spectral resolutions: dash-dotted lines correspond to filters with FWHM = 0.5 nm centered at 430.4 nm and 430.5 nm, while dashed lines correspond to filters with FWHM = 0.01 nm centered at the same wavelengths. These profiles illustrate the effect of spectral broadening on the observed G-band.}
    \label{fig:spatial averaged spectrum}
\end{figure*}


Figure~\ref{fig:spatial averaged spectrum} presents the spatially averaged spectra for the QS simulation in the G-band wavelength range (430-431 nm).
The black solid curve corresponds to the synthetic spectrum, while the red solid line shows the ATLAS spectrum\footnote{\url{https://bass2000.obspm.fr/solar_spect.php}}.
The dashed curve represents the narrowband Gaussian filter with FWHM = 0.01 nm, whereas the dashed line shows the wideband Gaussian filter with FWHM = 0.5 nm for spectral convolution applied to simulate instrumental broadening and generate filtered intensity maps.
The synthetic spectrum shows good agreement with the ATLAS reference, confirming the realism of the simulated profiles.




\begin{figure*}
    \centering
    \includegraphics[width=\linewidth]{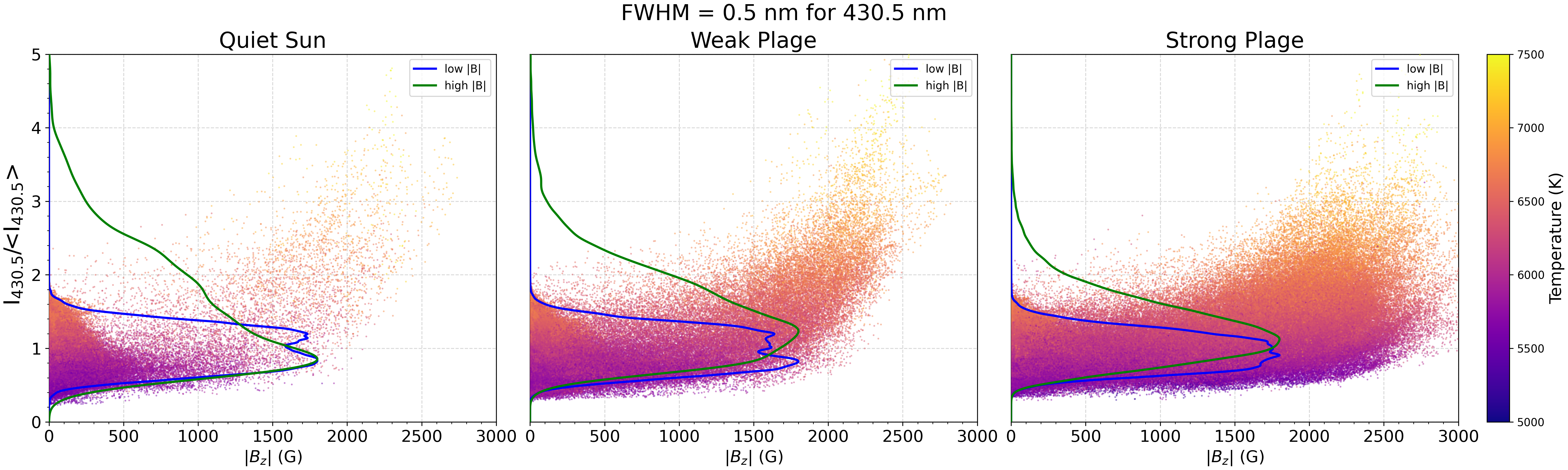}
    \includegraphics[width=\linewidth]{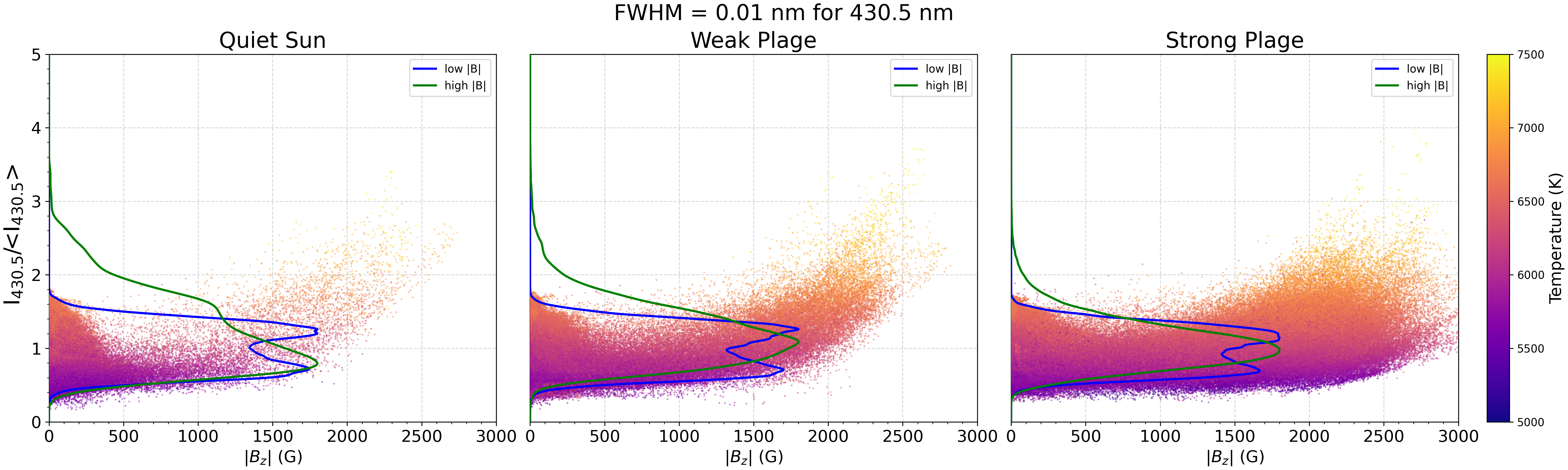}
    \caption{A six-panel scatter plot illustrating the relationship between the G-band brightness at 430.5 nm and the magnetic field strength at the optical depth layer $\tau_{500} = 1$. The top row corresponds to a Gaussian convolution with FWHM = 0.5 nm, while the bottom row corresponds to FWHM = 0.01 nm. Each column represents a different simulation setup with average magnetic field strengths of 10 G, 50 G, and 200 G. Data points are color-coded by temperature at $\tau_{500} = 1$. Distributions are shown separately for regions with $B \geq 1000$ G and $B < 1000$ G.}
    \label{fig:scatter magnetic field 4305 fwhms}
\end{figure*}

\begin{figure*}
    \centering
    \includegraphics[width=\linewidth]{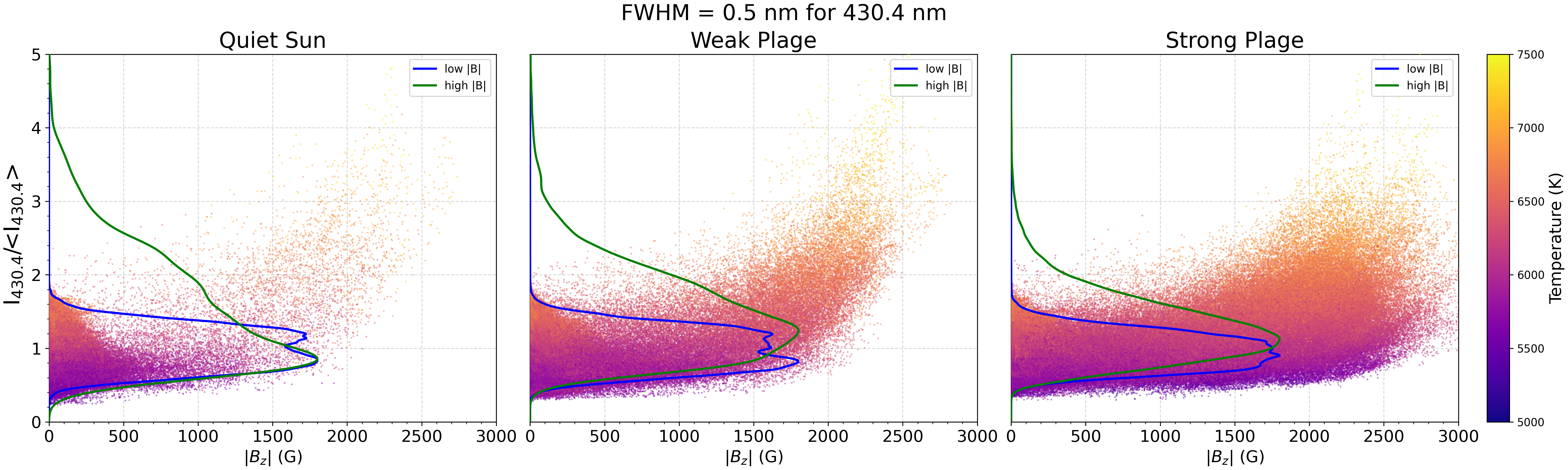}
    \includegraphics[width=\linewidth]{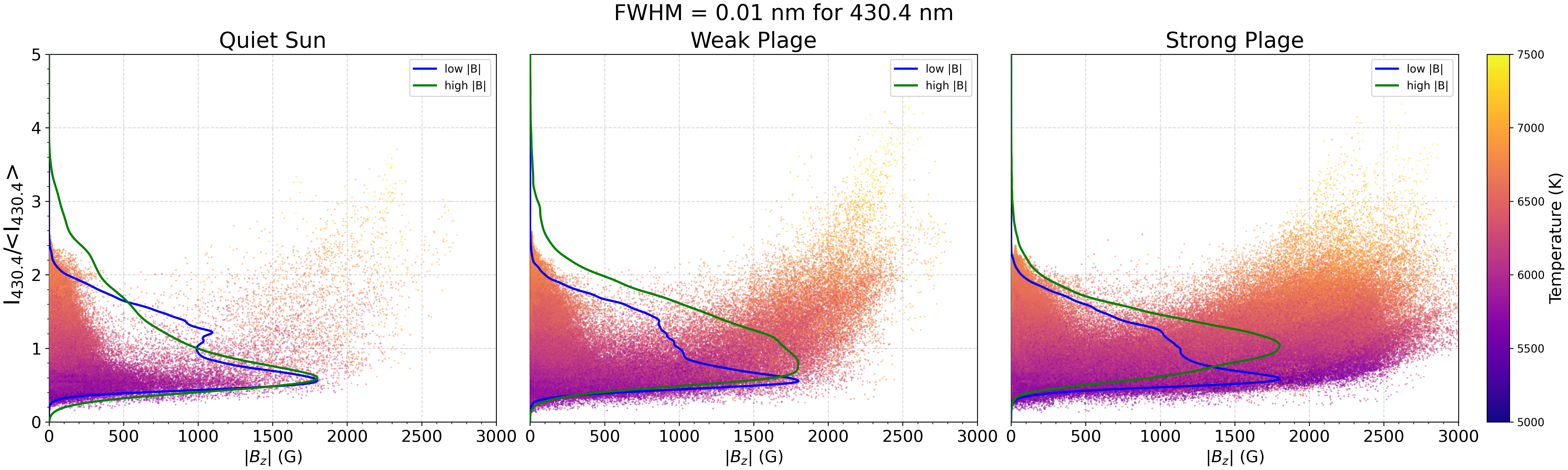}
    \caption{A six-panel scatter plot illustrating the relationship between the G-band brightness at 430.4 nm and the magnetic field strength at the optical depth layer $\tau_{500} = 1$. The top row corresponds to a Gaussian convolution with FWHM = 0.5 nm, while the bottom row corresponds to FWHM = 0.01 nm. Each column represents a different simulation setup with average magnetic field strengths of 10 G, 50 G, and 200 G. Data points are color-coded by temperature at $\tau_{500} = 1$. Distributions are shown separately for regions with $B \geq 1000$ G and $B < 1000$ G.}
    \label{fig:scatter magnetic field 4304 fwhms}
\end{figure*}


\begin{figure*}
    \centering
    \includegraphics[width=\linewidth]{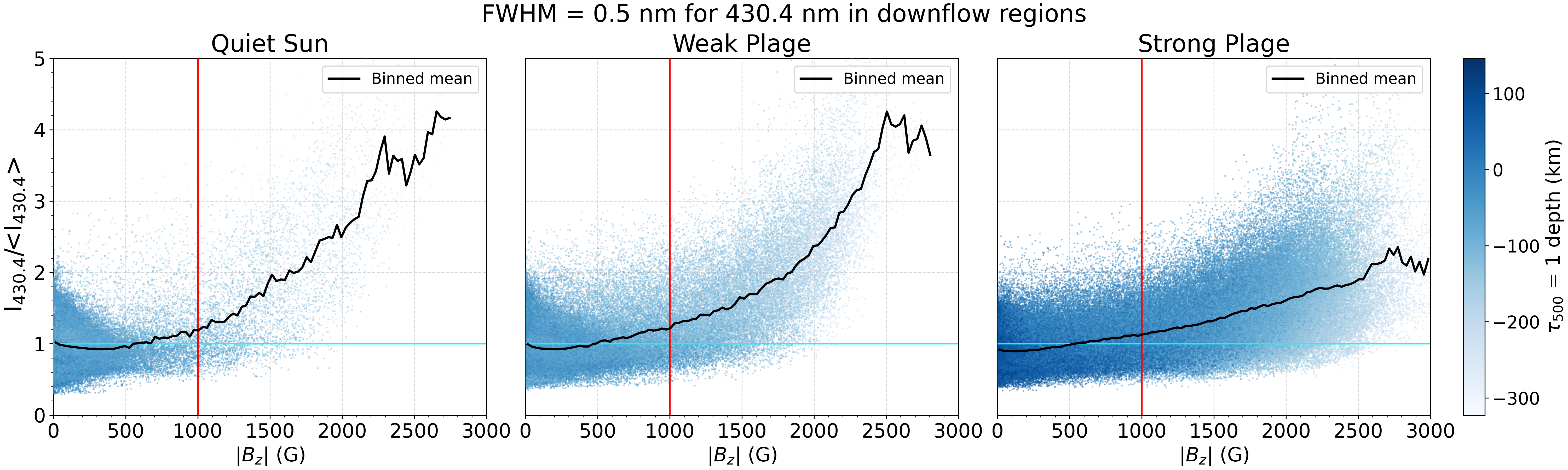}
    \includegraphics[width=\linewidth]{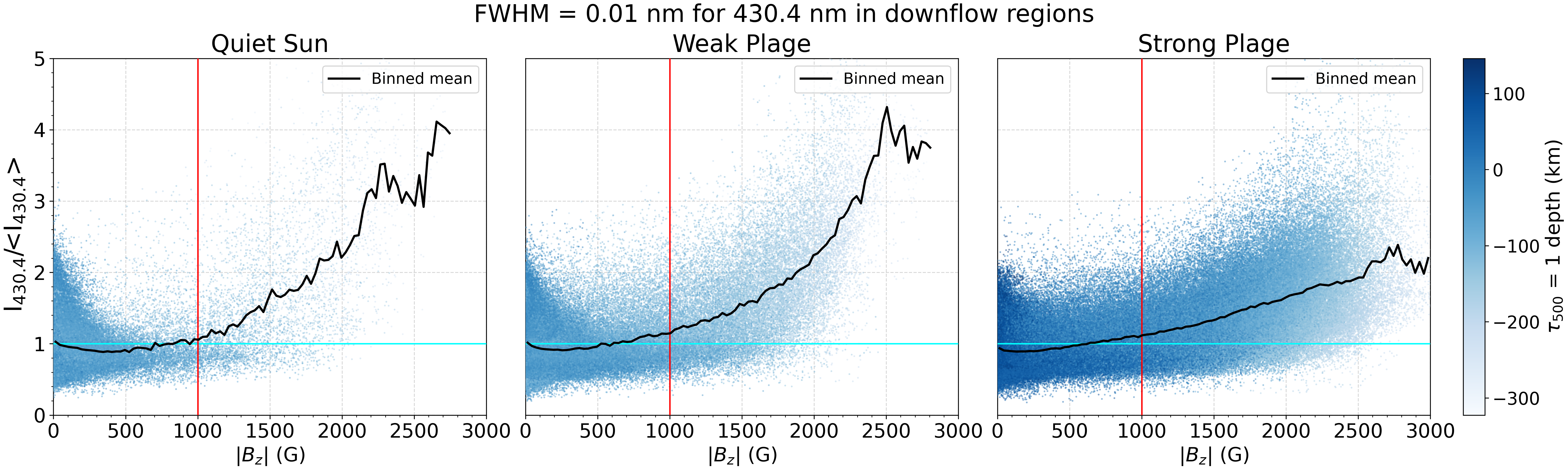}
    \caption{A six-panel scatter plot illustrating the relationship between the G-band brightness at 430.4 nm with the magnetic field strength at the optical depth layer $\tau_{500} = 1$ only including downflow regions ($v_z<0$). The different rows from top to bottom show the convolution of data with different FWHMs = 0.5 nm (top row) and 0.01 nm (bottom row). Each column corresponds to a different simulation setup. The data points in the scatter plot are colour-coded to represent the depression of $\tau_{500} = 1$ layer in comparison to z = 0 layer.}
    \label{fig:scatter magnetic field 4304 fwhms masked}
\end{figure*}



Fig. \ref{fig:scatter magnetic field 4305 fwhms} 
presents scatter plots showing the relationship between the G-band brightness and the vertical magnetic field strength (|$B_z$|) at the optical depth layer $\tau_{500} = 1$, for the filter centered at 430.5 nm.
The top row corresponds to the broadband with FWHM = 0.5 nm, while the bottom row corresponds to the narrowband with FWHM = 0.01 nm.
The three columns represent the QS, WP, and SP simulation setups, respectively. 
The green and blue lines represent density contours of the scatter distribution, highlighting regions with higher point concentration and emphasizing the spread of brightness values across different magnetic field strengths.
Individual points are color-coded according to the temperature at $\tau_{500} = 1$, allowing a direct connection between thermal structure and brightness response.
Figure \ref{fig:scatter magnetic field 4304 fwhms} is a similar figure, but for the filters centered at 430.4 nm.

From both figures, we find that G-band brightness increases with magnetic field strength, reflecting the partial evacuation of flux tubes and consequent dissociation of CH molecules.
The comparison between the two wavelengths highlights that for the broadband case (FWHM = 0.5 nm), the integrated contribution across the wider spectral range smooths out differences between 430.5 nm and 430.4 nm.
Because the two central wavelengths differ by only 0.1 nm, the broad filters capture similar average responses to magnetic fields.
However, at narrower bandwidths (FWHM = 0.01 nm), distinct differences in contrast emerge, with much higher intensity values in the 430.4 nm cases.
This demonstrates that the choice of central wavelength becomes critical when narrowband filters are employed. Although such bandwidths are comparable to the intrinsic line widths and therefore highly sensitive to the precise central wavelength, this sensitivity itself provides insight into how finely tuned the spectral sampling must be for reliable magnetic diagnostics at this wavelength.


As the filter width decreases to 0.01 nm, the intensity contrast reduces overall, with a more pronounced decrease for 430.5 nm in the SP region. This is evident in Fig. \ref{fig:scatter magnetic field 4305 fwhms}, where the scatter points become less dispersed and cluster more tightly at the narrow FWHM. Regions with strong magnetic fields generally exhibit enhanced brightness due to higher temperatures. However, some strong-field regions show lower-than-average brightness, indicating that the relationship between brightness and field strength is not always straightforward. The green and blue density contours in the scatter plots highlight this behavior, showing a broader brightness distribution in strong-field regions compared to weak-field areas.

In the QS simulation, the weaker and more diffuse magnetic fields lead to reduced brightness and less pronounced correlations in the scatter plots. In contrast, the Weak and SP setups display stronger, more concentrated fields, which produce larger brightness and temperature variations. Interestingly, in both the WP and SP regions, a fraction of points correspond to high magnetic field strengths but do not show the expected brightness enhancement, instead falling below the mean intensity line ($I / \langle I \rangle = 1$).

A key difference emerges when comparing filter widths and wavelength centers: for FWHM = 0.01 nm, the intensity contrast is stronger for 430.4 nm, whereas at 430.5 nm the contrast diminishes. By comparison, the broader 0.5 nm filter averages over nearby CH transitions, reducing the differences between the two central wavelengths and producing a more uniform response.

To understand the origin of reduced brightness in strongly magnetized regions, we examine the Wilson depression depth (i.e., the geometric displacement of the $\tau_{500} = 1$) as shown in Figure~\ref{fig:scatter magnetic field 4304 fwhms masked}. This figure presents the scatter plot between the G-band intensity at 430.4 nm (with FWHM = 0.01 nm) and the vertical magnetic field strength, similar to Figure~\ref{fig:scatter magnetic field 4304 fwhms}, but restricted to downflow regions. By filtering out upflow-associated pixels, we isolate the locations where magnetic bright points predominantly occur.

Even after masking granules, the mean trend (black line) retains the characteristic `fishhook' structure, consistent with earlier studies of magneto-convection \citep{Schnerr_2011, criscouli_uitenbroek_2014}.
In this context, the brightness initially decreases at weak field strengths because intergranular downflows without strong magnetic concentrations are relatively dark.
As the magnetic field intensifies, flux tubes within these downflows become partially evacuated, lowering gas pressure and deepening the Wilson depression. 
This enables radiation to escape from deeper, hotter layers, which leads to a recovery of brightness in the strong-field regime \citep{Spruit_1976, Schussler_2003}.


Notably, some strong-field regions, particularly in plage, do not exhibit the expected brightness enhancements.
Instead, they show a depression in the $\tau_{500} = 1$ layer and reduced intensity, consistent with the results of \citet{Riethmuller_2017}.
This darkening arises because radiative heating from the surrounding hot walls is sometimes insufficient to compensate for the suppressed convective energy transport within the cores of large flux patches.
While the edges of these magnetic concentrations can still benefit from lateral heating and Wilson depression effects, their highly evacuated cores remain comparatively dark. 
This contrast highlights a key difference between compact magnetic elements, such as bright points that sustain enhanced brightness, and extended flux patches in plage, where inner regions often display significant brightness deficits.
Additionally, bright points are not exclusively magnetic in origin. Observations have shown that localized brightness enhancements can also arise in weak-field regions, often referred to as non-magnetic bright points (NMBPs) \citep{Berger_Title_nmbp_2001, Langhans_nmbp_2002}. 
Simulations by \citet{Calvo_nmbp_2016} further demonstrated that such features can form through localized density depressions and vortex-like downdrafts, which create vertically extended structures.
These reduce the mass density and allow radiation to escape from deeper, hotter layers, producing bright point signatures without requiring strong magnetic fields.


Since the continuum primarily samples deeper layers than those corresponding to 430.4 nm, where temperature variations are smaller, its intensity distribution exhibits lower contrast. In contrast, the enhanced intensity contrast at 430.4 nm arises because its $\tau = 1$ surface forms higher in the atmosphere compared to the continuum. This elevated formation height results in greater intensity differences between granules and intergranular lanes. Such increased contrast is particularly useful in regions with small-scale temperature fluctuations or where the $\tau = 1$ surface spans a wider range of temperatures.

\begin{figure*}
    \centering
    \includegraphics[width=\linewidth]{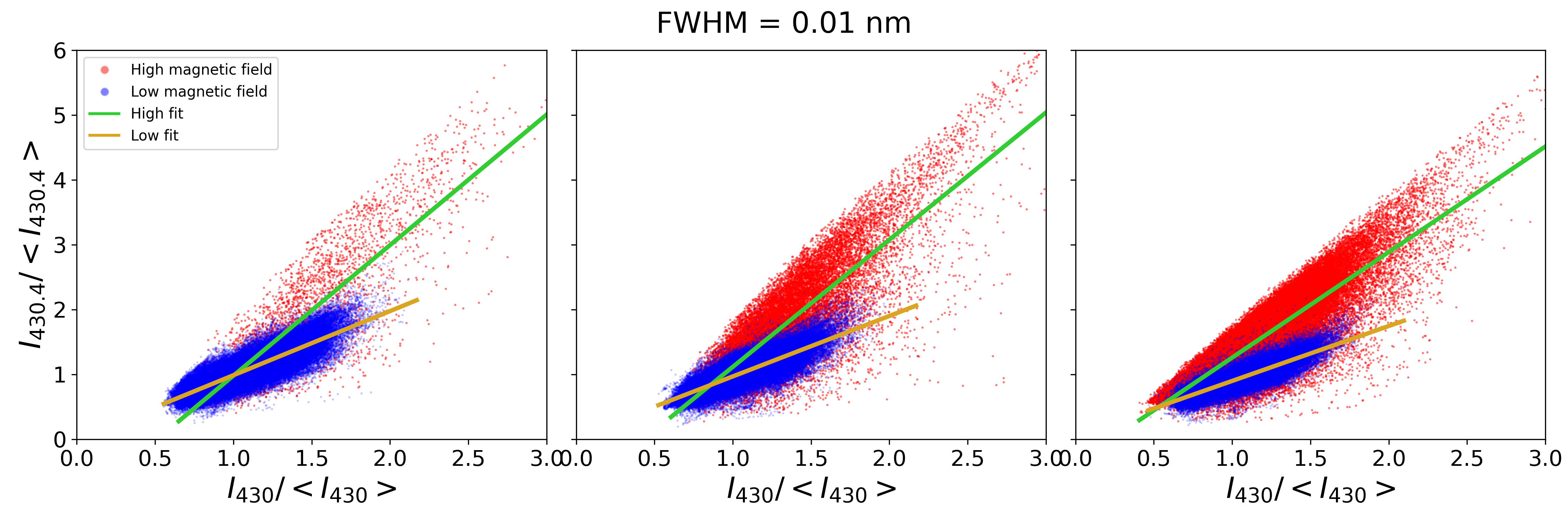}
    \includegraphics[width = \linewidth]{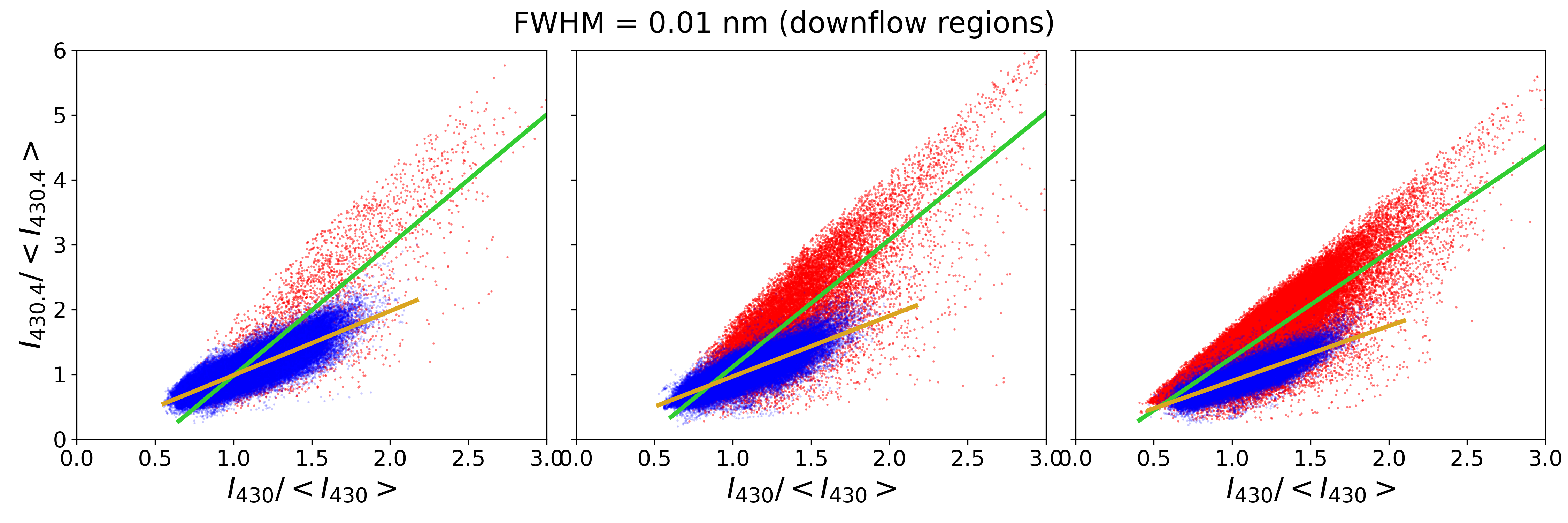}
    \caption{Scatter plots showing the relationship between continuum intensity and the G-band intensity at 430.4 nm for a filter width of 0.01 nm. The top row presents the whole domain, while the bottom row includes only downflow regions ($v_z < 0$), corresponding to intergranular lanes where magnetic bright points typically form. Columns corresponds to three different simulations setups. Red points indicate strong magnetic field regions ($|B| \geq 1000G$), while blue points indicate weak field regions ($|B| < 10 G$). For clarity, linear fits are overplotted: green lines correspond to strong-field regions and yellow lines to weak field regions. The slopes of the respective fits are provided in the Table~\ref{tab:slopes}.}
    \label{fig:scatter continuum fwhm = 0.01}
\end{figure*}

\begin{figure*}
    \centering
    \includegraphics[width=1\linewidth]{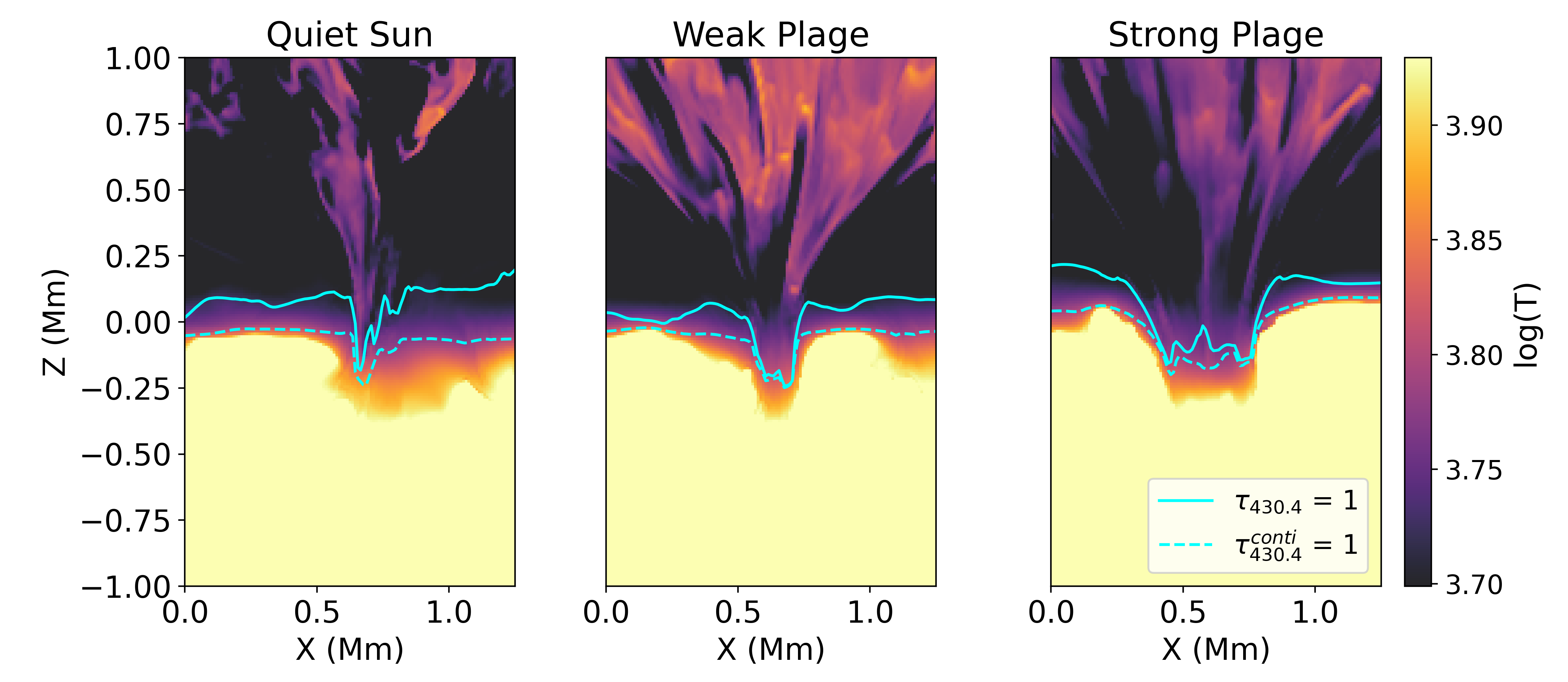}
    \caption{Vertical slices of the 3D simulation domain for a single magnetic feature across the three different magnetic field setups for 10 Gauss, 50 Gauss and 200 Gauss from left to right, illustrating the $\tau = 1$ surfaces for 430.4 nm (solid line) and the continuum at 430.4 nm (dashed line). The background color map represents the temperature distribution inside the magnetic features.}
    \label{fig:temp_slice}
\end{figure*}

Figure \ref{fig:scatter continuum fwhm = 0.01} shows scatter plots comparing the continuum intensity with the G-band brightness at 430.4 nm.
The top row corresponds to the full simulation domains, while the bottom row shows the same relation after masking out the upflow regions ($v_z < 0$).
The three columns represent the QS, WP and SP simulation setups.
The scatter points are classified by magnetic field strength, with red indicating strongly magnetized regions (|B| $\geq$ 1000 G) and blue indicating weakly magnetized regions (|B| < 10 G).
To highlight the overall trends, linear fits are overplotted: green lines correspond to strong-field regions and yellow lines to weak field regions.

The fitted slopes of these linear trends are summarized in Table~\ref{tab:slopes}.
A clear pattern emerges across all simulation setups: strong-field regions consistently follow steeper slopes than weak-field regions, and the slopes become even steeper when only downflow regions are considered.
In particular, the QS and WP cases show strong-field slopes exceeding 2.0 and 1.9, respectively, after masking, compared to significantly shallower slopes in the unmasked case.
Even in the SP simulation, where both strong and weak fields show reduced sensitivity, the strong-field slopes remain noticeably enhanced relative to the unmasked results.
These findings confirm that 430.4 nm brightness responds more sensitively to continuum intensity variations, especially in downflow regions where magnetic concentrations are most prominent.

\begin{table}
\centering
\caption{Slopes of the linear fits between continuum intensity at 430 nm and G-band intensity at 430.4 nm. 
Values are shown for strong-field regions ($|B| \geq 1000$ G) and weak-field regions ($|B| < 10$ G), 
for both unmasked and masked cases ($v_z < 0$).}
\label{tab:slopes}
\begin{tabular}{lcccc}
\hline
 & \multicolumn{2}{c}{full domain} & \multicolumn{2}{c}{only downflows ($v_z < 0$)} \\
Region & Strong field & Weak field & Strong field & Weak field \\
\hline
\textbf{QS}   & 1.775 & 1.310 & 2.013 & 0.989 \\
\textbf{WP}  & 1.810 & 1.330 & 1.962 & 0.933 \\
\textbf{SP} & 1.565 & 1.229 & 1.628 & 0.847 \\
\hline
\end{tabular}
\end{table}

The QS exhibits a relatively uniform distribution of normalized intensities ($I_{430.4} / \langle I_{430.4} \rangle$ vs $I_430 / \langle I_430 \rangle$), with strongly magnetized regions (red points) following steeper trends than weak-field regions (blue points), consistent with a greater enhancement of G-band brightness in the presence of strong magnetic fields. 
However, not all magnetic elements appear as bright points, since efficient heating—either laterally from surrounding granules or vertically from deeper layers—is required to produce significant brightness \citep{Ishikawa_2007}. 
In weak-field regions this heating is often insufficient, leading to cases where magnetic fields exist without corresponding brightness enhancements.  

Masking out upflows isolates regions where magnetic flux tubes typically reside. In this case, the weak-field distribution narrows, while strong-field points remain aligned along steeper slopes, emphasizing the central role of flux concentrations in producing enhanced G-band brightness. 
This effect is especially clear in the plage simulations, where strong-field regions dominate the intergranular lanes and the contrast between weak and strong fields becomes more pronounced.

To further investigate the contrast variations observed in Fig. \ref{fig:different intensities} and the differences between the intensity maps for continuum and 430.4 nm, Fig. \ref{fig:temp_slice} presents vertical temperature slices for three selected magnetic features in the QS, WP, and SP regions. The background color represents the temperature distribution, while the overlaid curves indicate the $\tau = 1$ layers for 430.4 nm (solid line) and the continuum at 430.4 nm (dashed line). This figure reveals that the formation height of the 430.4 nm intensity is consistently higher than that of the nearby continuum, spanning a broader range of temperature variations.
As a result, intensity maps at 430.4 nm show increased contrast because, over granules, they form in relatively cooler layers compared to the continuum, whereas in the intergranular lanes, they originate at similar depths.

Interpreting G-band brightness is therefore non-trivial, as it depends not only on the intrinsic properties of flux tubes but also on observational effects such as spatial resolution, thresholding, and alignment accuracy, all of which can alter the apparent correspondence between magnetic field strength and bright features \citep{criscouli_uitenbroek_2014}.

\section{Discussions}
\label{sec:discussions}
In this study, we explored the diagnostic capabilities of the G-band at 430.4\,nm for detecting small-scale magnetic structures in the solar photosphere. 
Using three-dimensional radiative MHD simulations from the MURaM code and synthetic spectra generated with RH 1.5D code, we analysed the intensity contrast  of the G-band under varying magnetic field conditions for different wavelength centers with narrowband filters.
The observed brightness enhancements reflect the combined effect of molecular opacity variations and magnetic structuring of the photosphere, as discussed in detail in the previous sections.


A key focus of this work was the comparative analysis of intensity contrast at 430.4\,nm and the more commonly used 430.5\,nm wavelength, across different filter widths.
While both wavelengths are sensitive to underlying magnetic structures, we find that the contrast at 430.5\,nm degrades significantly when the spectral filter width is reduced from 0.5\,nm to 0.01\,nm.


A 0.01 nm passband therefore does not represent a standard present-day or near-future spectral resolution for G-band observations. Rather, it is used here as a diagnostic tool to isolate and compare the contributions from very narrow spectral intervals around 430.4 nm and 430.5 nm. The effects identified at these narrow intervals are subsequently integrated within the broader passbands employed by current instruments and those anticipated for future G-band observations.


The motivation for this narrow-band analysis is thus to propose 430.4 nm as the optimal choice of the central wavelength for the G-band filters. By resolving the spectral structure at high resolution, we demonstrate that the 430.4 nm region is intrinsically more sensitive to magnetic and thermal variations than 430.5 nm. As a result, even when employing realistic, currently available or near-future G-band filters, positioning the passband closer to 430.4 nm rather than 430.5 nm is expected to produce higher contrast, because the enhanced spectral sensitivity of the CH features at 430.4 nm is partially retained after convolution with broader filters.


Based on our findings, the 430.4\,nm region emerges as a promising alternative to the traditional 430.5\,nm G-band center within the assumptions of the present modeling. The enhanced contrast observed at this wavelength highlights its diagnostic potential for narrowband imaging of magnetic bright points. However, this conclusion is based on idealized numerical experiments and should be tested further by including realistic instrumental degradation and, whenever possible, through dedicated observations.


We also confirm that not all bright points detected in the G-band are of magnetic origin.
Localized brightness enhancements can occur in weak-field regions through non-magnetic processes such as density depressions and vortex-like motions, which expose deeper layers and increase radiative output.
This result highlights the need for caution when interpreting G-band observations as direct proxies for magnetism.

In addition, our study emphasizes the complex interplay between magnetic field strength, CH molecule dissociation, and radiative transfer effects in shaping G-band intensity variations.
The occurrence of large flux patches that appear dim despite hosting strong magnetic fields points to the influence of flux tube geometry, reduced heating efficiency, and opacity effects.
These findings underscore the importance of high-resolution spectral synthesis and imaging strategies that account for the molecular properties of CH transitions.


We next consider the implications of using 1.5D radiative transfer. Horizontal radiative transfer can be important for spectral lines whose source function are weakly coupled to the local thermodynamic conditions. \citet{Leenaarts_2012} showed that lines such as H$\alpha$, which have very low photon destruction probabilities and are dominated by scattering, full 3D radiative transfer is esential because lateral photon diffusion strongly influences the emergent intensity. In contrast, CH molecular lines form in the deep photopshere under conditions close to LTE, where collisional rates are high and the source function is tightly coupled to the local temperature. In this regime, the photon destruction probability is large and the mean free path of photons is short, so horizontal radiative transfer plays a much smaller role in shaping the emergent intensity. Consequently, the G-band signal is controlled primarily by the local atmospheric structure rather than by lateral radiative coupling. The use of 1.5D synthesis therefore represents a well-established and physically justified approximation that captures the dominant physics of CH line formation while enabling a statistically meaningful and computationally feasible analysis of the simulation domain.


We note that the present analysis does not include instrumental spatial degradation or noise, and therefore represents an optimistic estimate of the achievable contrast; future work will incorporate full instrumental point-spread functions while spectral degradation has already been accounted for through Gaussian filtering.

Future work will extend this study by examining the magnetic diagnostic potential of the 430.4 nm region through detailed spectropolarimetric synthesis.
Building on the intensity contrast analysis presented here, we will incorporate the full set of Stokes parameters to evaluate how magnetic fields imprint themselves on molecular and atomic lines within this wavelength range.
Particular emphasis will be placed on the Stokes V signal, which directly probes the LOS component of the magnetic field through circular polarization.
By quantifying the sensitivity of the Stokes V profiles at 430.4 nm to different magnetic field strengths, we aim to determine the extent to which this wavelength can provide reliable measurements of photospheric magnetism.

This extension is especially relevant because traditional G-band studies centered at 430.5 nm have largely been limited to intensity diagnostics, leaving their spectropolarimetric capabilities less explored. 
Demonstrating that the 430.4 nm region can deliver both brightness and magnetic information would not only broaden the utility of the G-band observations but also provide a valuable alternative in situations where simultaneous imaging and magnetic field retrieval are required.
In this context, narrowband imaging techniques could be adapted to extract polarization signatures, offering practical pathways for observational setups where conventional spectropolarimetry is challenging.

\section*{Data Availability}
Data sets generated during the current study are available from the corresponding author on a reasonable request. 


\section*{Acknowledgements}
S.P. acknowledges support from the CEFIPRA Research Project No. 6904-2. 
N.Y. acknowledges the support from the DST INSPIRE Faculty Grant (IF21-PH-268) and the SERB MATRICS grant (MTR/2023/001332).




\bibliographystyle{mnras}
\bibliography{example} 

@article{Lites_2009,
       author = {{Lites}, B.~W.},
        title = "{The Topology and Behavior of Magnetic Fields Emerging at the Solar Photosphere}",
      journal = {\ssr},
         year = 2009,
        month = apr,
       volume = {144},
       number = {1-4},
        pages = {197-212},
          doi = {10.1007/s11214-008-9437-x},
       adsurl = {https://ui.adsabs.harvard.edu/abs/2009SSRv..144..197L}
}

@article{Leenaarts_2012,
doi = {10.1088/0004-637X/749/2/136},
url = {https://doi.org/10.1088/0004-637X/749/2/136},
year = {2012},
month = {apr},
publisher = {The American Astronomical Society},
volume = {749},
number = {2},
pages = {136},
author = {Leenaarts, J. and Carlsson, M. and Rouppe van der Voort, L.},
title = {THE FORMATION OF THE Hα LINE IN THE SOLAR CHROMOSPHERE},
journal = {The Astrophysical Journal}
}

@article{de_Wijn_2008,
   title={HinodeObservations of Magnetic Elements in Internetwork Areas},
   volume={684},
   ISSN={1538-4357},
   url={http://dx.doi.org/10.1086/590237},
   DOI={10.1086/590237},
   number={2},
   journal={The Astrophysical Journal},
   publisher={American Astronomical Society},
   author={de Wijn, A. G. and Lites, B. W. and Berger, T. E. and Frank, Z. A. and Tarbell, T. D. and Ishikawa, R.},
   year={2008},
   month=sep, pages={1469–1476} }

@article{Schussler_2003,
  author  = {Schüssler, M. and Shelyag, S. and Berdyugina, S. and Vögler, A. and Solanki, S. K.},
  title   = {Why Solar Magnetic Flux Concentrations Are Bright in Molecular Bands},
  journal = {ApJ},
  year    = {2003},
  month   = {nov},
  volume  = {597},
  number  = {2},
  pages   = {L173--L176},
  doi     = {10.1086/379869},
  url     = {https://ui.adsabs.harvard.edu/abs/2003ApJ...597L.173S}
}

@ARTICLE{Shelyag_2004,
       author = {{Shelyag}, S. and {Sch{\"u}ssler}, M. and {Solanki}, S.~K. and {Berdyugina}, S.~V. and {V{\"o}gler}, A.},
        title = "{G-band spectral synthesis and diagnostics of simulated solar magneto-convection}",
      journal = {\aap},
         year = 2004,
        month = nov,
       volume = {427},
        pages = {335-343},
          doi = {10.1051/0004-6361:20040471},
       adsurl = {https://ui.adsabs.harvard.edu/abs/2004A&A...427..335S}
}

@ARTICLE{Sanchez_2004,
       author = {{S{\'a}nchez Almeida}, J. and {M{\'a}rquez}, I. and {Bonet}, J.~A. and {Dom{\'\i}nguez Cerde{\~n}a}, I. and {Muller}, R.},
        title = "{Bright Points in the Internetwork Quiet Sun}",
      journal = {\apjl},
         year = 2004,
        month = jul,
       volume = {609},
       number = {2},
        pages = {L91-L94},
          doi = {10.1086/422752},
archivePrefix = {arXiv},
       eprint = {astro-ph/0405515},
 primaryClass = {astro-ph},
       adsurl = {https://ui.adsabs.harvard.edu/abs/2004ApJ...609L..91S}
}

@ARTICLE{Muller_Roudier_1984,
  title    = "Variability of the quiet photospheric network",
  author   = "Muller, R and Roudier, Th",
  journal  = "Solar Physics",
  volume   =  94,
  number   =  1,
  pages    = "33--47",
  month    =  aug,
  year     =  1984
}

@ARTICLE{Berger_Title_1995,
       author = {{Berger}, T.~E. and {Schrijver}, C.~J. and {Shine}, R.~A. and {Tarbell}, T.~D. and {Title}, A.~M. and {Scharmer}, G.},
        title = "{New Observations of Subarcsecond Photospheric Bright Points}",
      journal = {\apj},
         year = 1995,
        month = nov,
       volume = {454},
        pages = {531},
          doi = {10.1086/176504},
       adsurl = {https://ui.adsabs.harvard.edu/abs/1995ApJ...454..531B}
}

@ARTICLE{Schussler_Vogler_2003,
       author = {{V{\"o}gler}, A. and {Sch{\"u}ssler}, M.},
        title = "{Studying magneto-convection by numerical simulation}",
      journal = {Astronomische Nachrichten},
         year = 2003,
        month = jan,
       volume = {324},
       number = {4},
        pages = {399-404},
          doi = {10.1002/asna.200310146},
       adsurl = {https://ui.adsabs.harvard.edu/abs/2003AN....324..399V}
}

@article{Uitenbroek_2004,
doi = {10.1086/382037},
url = {https://dx.doi.org/10.1086/382037},
year = {2004},
month = {apr},
publisher = {},
volume = {604},
number = {2},
pages = {960},
author = {Uitenbroek, H. and Miller-Ricci, E. and Ramos, A. Asensio and Bueno, J. Trujillo},
title = {The Zeeman Effect in the G Band},
journal = {The Astrophysical Journal}
}

@article{Vogler_2005,
  author  = {Vögler, A. and Shelyag, S. and Schüssler, M. and Cattaneo, F. and Emonet, T. and Linde, T.},
  title   = {Simulations of magneto-convection in the solar photosphere -- Equations, methods, and results of the MURaM code},
  journal = {A\&A},
  year    = {2005},
  volume  = {429},
  number  = {1},
  pages   = {335--351},
  doi     = {10.1051/0004-6361:20041507},
  url     = {https://doi.org/10.1051/0004-6361:20041507}
}

@ARTICLE{Uitenbroek_RH_2001,
       author = {{Uitenbroek}, H.},
        title = "{Multilevel Radiative Transfer with Partial Frequency Redistribution}",
      journal = {\apj},
         year = 2001,
        month = aug,
       volume = {557},
       number = {1},
        pages = {389-398},
          doi = {10.1086/321659},
       adsurl = {https://ui.adsabs.harvard.edu/abs/2001ApJ...557..389U}
}

@ARTICLE{Tiago_Periera_RH_2015,
       author = {{Pereira}, Tiago M.~D. and {Uitenbroek}, Han},
        title = "{RH 1.5D: a massively parallel code for multi-level radiative transfer with partial frequency redistribution and Zeeman polarisation}",
      journal = {\aap},
         year = 2015,
        month = feb,
       volume = {574},
          eid = {A3},
        pages = {A3},
          doi = {10.1051/0004-6361/201424785},
archivePrefix = {arXiv},
       eprint = {1411.1079},
 primaryClass = {astro-ph.SR},
       adsurl = {https://ui.adsabs.harvard.edu/abs/2015A&A...574A...3P}
}

@INPROCEEDINGS{steiner_bruls_2001,
       author = {{Steiner}, O. and {Bruls}, J. and {Hauschildt}, P.~H.},
        title = "{Why are G-Band Bright Points Bright?}",
    booktitle = {Advanced Solar Polarimetry -- Theory, Observation, and Instrumentation},
         year = 2001,
       editor = {{Sigwarth}, Michael},
       series = {Astronomical Society of the Pacific Conference Series},
       volume = {236},
        month = jan,
        pages = {453},
       adsurl = {https://ui.adsabs.harvard.edu/abs/2001ASPC..236..453S}
}

@ARTICLE{Spruit_1976,
       author = {{Spruit}, H.~C.},
        title = "{Pressure equilibrium and energy balance of small photospheric fluxtubes.}",
      journal = {\solphys},
         year = 1976,
        month = nov,
       volume = {50},
       number = {2},
        pages = {269-295},
          doi = {10.1007/BF00155292},
       adsurl = {https://ui.adsabs.harvard.edu/abs/1976SoPh...50..269S}
}

@article{Carlsson_2004,
doi = {10.1086/423305},
url = {https://dx.doi.org/10.1086/423305},
year = {2004},
month = {jun},
publisher = {},
volume = {610},
number = {2},
pages = {L137},
author = {Carlsson, Mats and Stein, Robert F. and Nordlund, {\AA}ke and Scharmer, Göran B.},
title = {Observational Manifestations of Solar Magnetoconvection: Center-to-Limb Variation},
journal = {The Astrophysical Journal}
}

@article{Almeida_2001,
   title={G‐Band Spectral Synthesis in Solar Magnetic Concentrations},
   volume={555},
   ISSN={1538-4357},
   url={http://dx.doi.org/10.1086/321521},
   DOI={10.1086/321521},
   number={2},
   journal={The Astrophysical Journal},
   publisher={American Astronomical Society},
   author={Almeida, J. Sanchez and Ramos, A. Asensio and Bueno, J. Trujillo and Cernicharo, J.},
   year={2001},
   month=jul, pages={978–989} }

@ARTICLE{Uitenbroek_Trits_2006,
       author = {{Uitenbroek}, H. and {Tritschler}, A.},
        title = "{The Contrast of Magnetic Elements in Synthetic CH- and CN-Band Images of Solar Magnetoconvection}",
      journal = {\apj},
         year = 2006,
        month = mar,
       volume = {639},
       number = {1},
        pages = {525-533},
          doi = {10.1086/499331},
archivePrefix = {arXiv},
       eprint = {astro-ph/0510333},
 primaryClass = {astro-ph},
       adsurl = {https://ui.adsabs.harvard.edu/abs/2006ApJ...639..525U}
}

@ARTICLE{Knoelker_1991,
       author = {{Knoelker}, M. and {Grossmann-Doerth}, U. and {Schuessler}, M. and {Weisshaar}, E.},
        title = "{Some developments in the theory of magnetic flux concentrations in the solar atmosphere}",
      journal = {Advances in Space Research},
         year = 1991,
        month = jan,
       volume = {11},
       number = {5},
        pages = {285-295},
          doi = {10.1016/0273-1177(91)90393-X},
       adsurl = {https://ui.adsabs.harvard.edu/abs/1991AdSpR..11e.285K}
}

@article{Stein_2006,
doi = {10.1086/501445},
url = {https://dx.doi.org/10.1086/501445},
year = {2006},
month = {may},
publisher = {},
volume = {642},
number = {2},
pages = {1246},
author = {Stein, R. F. and Nordlund, {\AA}.},
title = {Solar Small-Scale Magnetoconvection},
journal = {The Astrophysical Journal}
}

@article{Rempel_2014,
doi = {10.1088/0004-637X/789/2/132},
url = {https://dx.doi.org/10.1088/0004-637X/789/2/132},
year = {2014},
month = {jun},
publisher = {The American Astronomical Society},
volume = {789},
number = {2},
pages = {132},
author = {Rempel, M.},
title = {NUMERICAL SIMULATIONS OF QUIET SUN MAGNETISM: ON THE CONTRIBUTION FROM A SMALL-SCALE DYNAMO},
journal = {The Astrophysical Journal}
}

@ARTICLE{Deinzer_1984b,
       author = {{Deinzer}, W. and {Hensler}, G. and {Schussler}, M. and {Weisshaar}, E.},
        title = "{Model Calculations of Magnetic Flux Tubes - Part Two - Stationary Results for Solar Magnetic Elements}",
      journal = {\aap},
         year = 1984,
        month = oct,
       volume = {139},
        pages = {435},
       adsurl = {https://ui.adsabs.harvard.edu/abs/1984A&A...139..435D}
}

@ARTICLE{Deinzer_1984a,
       author = {{Deinzer}, W. and {Hensler}, G. and {Schuessler}, M. and {Weisshaar}, E.},
        title = "{Model calculations of magnetic flux tubes. I - Equations and method. II - Stationary results for solar magnetic elements}",
      journal = {\aap},
         year = 1984,
        month = oct,
       volume = {139},
       number = {2},
        pages = {426-449},
       adsurl = {https://ui.adsabs.harvard.edu/abs/1984A&A...139..426D}
}

@ARTICLE{Solanki_1993,
       author = {{Solanki}, Sami K.},
        title = "{Smallscale Solar Magnetic Fields - an Overview}",
      journal = {\ssr},
         year = 1993,
        month = mar,
       volume = {63},
       number = {1-2},
        pages = {1-188},
          doi = {10.1007/BF00749277},
       adsurl = {https://ui.adsabs.harvard.edu/abs/1993SSRv...63....1S}
}

@article{Riethmuller_2017,
  author  = {Riethmüller, T. L. and Solanki, S. K.},
  title   = {The dark side of solar photospheric G-band bright points},
  journal = {A\&A},
  year    = {2017},
  volume  = {598},
  pages   = {A123},
  doi     = {10.1051/0004-6361/201629773},
  url     = {https://doi.org/10.1051/0004-6361/201629773}
}

@article{Berger_1998,
doi = {10.1086/305309},
url = {https://dx.doi.org/10.1086/305309},
year = {1998},
month = {mar},
publisher = {},
volume = {495},
number = {2},
pages = {973},
author = {Berger, Thomas E. and Löfdahl, Mats G. and Shine, Richard S. and Title, Alan M.},
title = {Measurements of Solar Magnetic Element Motion from High-Resolution Filtergrams},
journal = {The Astrophysical Journal}
}

@article{schrijver1998,
  title={Large-scale coronal heating by the small-scale magnetic field of the Sun},
  author={Schrijver, Carolus J and Harvey, KL and Sheeley, NR and Wang, Y-M and Van den Oord, GHJ and Shine, RA and Tarbell, TD and Hurlburt, NE and others},
  journal={Nature},
  volume={394},
  number={6689},
  pages={152--154},
  year={1998},
  publisher={Nature Publishing Group}
}

@ARTICLE{Berdyugina_2003,
       author = {{Berdyugina}, S.~V. and {Solanki}, S.~K. and {Frutiger}, C.},
        title = "{The molecular Zeeman effect and diagnostics of solar and stellar magnetic fields. II. Synthetic Stokes profiles in the Zeeman regime}",
      journal = {\aap},
         year = 2003,
        month = dec,
       volume = {412},
        pages = {513-527},
          doi = {10.1051/0004-6361:20031473},
       adsurl = {https://ui.adsabs.harvard.edu/abs/2003A&A...412..513B}
}

@article{berdyugina_solanki_2002,
  author  = {Berdyugina, S. V. and Solanki, S. K.},
  title   = {The molecular Zeeman effect and diagnostics of solar and stellar magnetic fields -- I. Theoretical spectral patterns in the Zeeman regime},
  journal = {A\&A},
  year    = {2002},
  volume  = {385},
  number  = {2},
  pages   = {701--715},
  doi     = {10.1051/0004-6361:20020130},
  url     = {https://doi.org/10.1051/0004-6361:20020130}
}

@ARTICLE{Jorgensen_1996,
       author = {{Jorgensen}, U.~G. and {Larsson}, M. and {Iwamae}, A. and {Yu}, B.},
        title = "{Line intensities for CH and their application to stellar atmospheres.}",
      journal = {\aap},
         year = 1996,
        month = nov,
       volume = {315},
        pages = {204-211},
       adsurl = {https://ui.adsabs.harvard.edu/abs/1996A&A...315..204J}
}

@article{Beck_2007,
   title={Magnetic properties of G-band bright points in a sunspot moat},
   volume={472},
   ISSN={1432-0746},
   url={http://dx.doi.org/10.1051/0004-6361:20065620},
   DOI={10.1051/0004-6361:20065620},
   number={2},
   journal={Astronomy &amp; Astrophysics},
   publisher={EDP Sciences},
   author={Beck, C. and Bellot Rubio, L. R. and Schlichenmaier, R. and Sütterlin, P.},
   year={2007},
   month=jul, pages={607–622} }

@article{Ishikawa_2007,
   title={Relationships between magnetic foot points and G-band bright structures},
   volume={472},
   ISSN={1432-0746},
   url={http://dx.doi.org/10.1051/0004-6361:20066942},
   DOI={10.1051/0004-6361:20066942},
   number={3},
   journal={Astronomy &amp; Astrophysics},
   publisher={EDP Sciences},
   author={Ishikawa, R. and Tsuneta, S. and Kitakoshi, Y. and Katsukawa, Y. and Bonet, J. A. and Vargas Domínguez, S. and Rouppe van der Voort, L. H. M. and Sakamoto, Y. and Ebisuzaki, T.},
   year={2007},
   month=jul, pages={911–918} }

@article{Cheung_2008,
doi = {10.1086/591245},
url = {https://dx.doi.org/10.1086/591245},
year = {2008},
month = {nov},
publisher = {},
volume = {687},
number = {2},
pages = {1373},
author = {Cheung, M. C. M. and Schüssler, M. and Tarbell, T. D. and Title, A. M.},
title = {Solar Surface Emerging Flux Regions: A Comparative Study of Radiative MHD Modeling and Hinode SOT Observations},
journal = {The Astrophysical Journal}
}

@article{Schnerr_2011,
  author  = {Schnerr, R. S. and Spruit, H. C.},
  title   = {The brightness of magnetic field concentrations in the quiet Sun},
  journal = {A\&A},
  year    = {2011},
  volume  = {532},
  pages   = {A136},
  doi     = {10.1051/0004-6361/201015976},
  url     = {https://doi.org/10.1051/0004-6361/201015976}
}

@article{ Calvo_nmbp_2016,
	author = {Calvo, F. and Steiner, O. and Freytag, B.},
	title = {Non-magnetic photospheric bright points in 3D simulations  of the solar atmosphere⋆},
	DOI= "10.1051/0004-6361/201628649",
	url= "https://doi.org/10.1051/0004-6361/201628649",
	journal = {A&A},
	year = 2016,
	volume = 596,
	pages = "A43",
}

@article{ criscouli_uitenbroek_2014,
	author = {Criscuoli, S. and Uitenbroek, H.},
	title = {The statistical distribution of the magnetic-field strength in G-band bright points},
	DOI= "10.1051/0004-6361/201322909",
	url= "https://doi.org/10.1051/0004-6361/201322909",
	journal = {A&A},
	year = 2014,
	volume = 562,
	pages = "L1",
	month = "",
}

@article{Gošić_2014,
doi = {10.1088/0004-637X/797/1/49},
url = {https://dx.doi.org/10.1088/0004-637X/797/1/49},
year = {2014},
month = {nov},
publisher = {The American Astronomical Society},
volume = {797},
number = {1},
pages = {49},
author = {Gošić, M. and Rubio, L. R. Bellot and Orozco Suárez, D. and Katsukawa, Y. and del Toro Iniesta, J. C.},
title = {THE SOLAR INTERNETWORK. I. CONTRIBUTION TO THE NETWORK MAGNETIC FLUX},
journal = {The Astrophysical Journal}
}

@article{Buehler_2015,
   title={Properties of solar plage from a spatially coupled inversion of Hinode SP data},
   volume={576},
   ISSN={1432-0746},
   url={http://dx.doi.org/10.1051/0004-6361/201424970},
   DOI={10.1051/0004-6361/201424970},
   journal={Astronomy &amp; Astrophysics},
   publisher={EDP Sciences},
   author={Buehler, D. and Lagg, A. and Solanki, S. K. and van Noort, M.},
   year={2015},
   month=mar, pages={A27} }

@article{Orozco_Su_rez_2007,
   title={Quiet-Sun Internetwork Magnetic Fields from the Inversion of Hinode Measurements},
   volume={670},
   ISSN={1538-4357},
   url={http://dx.doi.org/10.1086/524139},
   DOI={10.1086/524139},
   number={1},
   journal={The Astrophysical Journal},
   publisher={American Astronomical Society},
   author={Orozco Suárez, D. and Bellot Rubio, L. R. and del Toro Iniesta, J. C. and Tsuneta, S. and Lites, B. W. and Ichimoto, K. and Katsukawa, Y. and Nagata, S. and Shimizu, T. and Shine, R. A. and Suematsu, Y. and Tarbell, T. D. and Title, A. M.},
   year={2007},
   month=oct, pages={L61–L64} 
}

@article{Ishikawa_2009,
   title={Comparison of transient horizontal magnetic fields in a plage region and in the quiet Sun},
   volume={495},
   ISSN={1432-0746},
   url={http://dx.doi.org/10.1051/0004-6361:200810636},
   DOI={10.1051/0004-6361:200810636},
   number={2},
   journal={Astronomy &amp; Astrophysics},
   publisher={EDP Sciences},
   author={Ishikawa, R. and Tsuneta, S.},
   year={2009},
   month=jan, pages={607–612} }

@article{Solanki_2006,
doi = {10.1088/0034-4885/69/3/R02},
url = {https://dx.doi.org/10.1088/0034-4885/69/3/R02},
year = {2006},
month = {feb},
publisher = {},
volume = {69},
number = {3},
pages = {563},
author = {Solanki, Sami K and Inhester, Bernd and Schüssler, Manfred},
title = {The solar magnetic field},
journal = {Reports on Progress in Physics}
}

@article{Hood_2011,
   title={Solar magnetic fields},
   volume={187},
   ISSN={0031-9201},
   url={http://dx.doi.org/10.1016/j.pepi.2011.04.010},
   DOI={10.1016/j.pepi.2011.04.010},
   number={3–4},
   journal={Physics of the Earth and Planetary Interiors},
   publisher={Elsevier BV},
   author={Hood, Alan W. and Hughes, David W.},
   year={2011},
   month=aug, pages={78–91} }

@Article{Scherrer_2012,
author={Scherrer, P. H.
and Schou, J.
and Bush, R. I.
and Kosovichev, A. G.
and Bogart, R. S.
and Hoeksema, J. T.
and Liu, Y.
and Duvall, T. L.
and Zhao, J.
and Title, A. M.
and Schrijver, C. J.
and Tarbell, T. D.
and Tomczyk, S.},
title={The Helioseismic and Magnetic Imager (HMI) Investigation for the Solar Dynamics Observatory (SDO)},
journal={Solar Physics},
year={2012},
month={Jan},
day={01},
volume={275},
number={1},
pages={207-227},
issn={1573-093X},
doi={10.1007/s11207-011-9834-2},
url={https://doi.org/10.1007/s11207-011-9834-2}
}

@article{Scharmer_2008,
doi = {10.1086/595744},
url = {https://dx.doi.org/10.1086/595744},
year = {2008},
month = {nov},
publisher = {},
volume = {689},
number = {1},
pages = {L69},
author = {Scharmer, G. B. and Narayan, G. and Hillberg, T. and de la Cruz Rodriguez, J. and Löfdahl, M. G. and Kiselman, D. and Sütterlin, P. and van Noort, M. and Lagg, A.},
title = {CRISP Spectropolarimetric Imaging of Penumbral Fine Structure},
journal = {The Astrophysical Journal}
}

@Article{Rimmele_2020,
author={Rimmele, Thomas R.
and Warner, Mark
and Keil, Stephen L.
and Goode, Philip R.
and Kn{\"o}lker, Michael
and Kuhn, Jeffrey R.
and Rosner, Robert R.
and McMullin, Joseph P.
and Casini, Roberto
and Lin, Haosheng
and W{\"o}ger, Friedrich
and von der L{\"u}he, Oskar
and Tritschler, Alexandra
and Davey, Alisdair
and de Wijn, Alfred
and Elmore, David F.
and Fehlmann, Andr{\'e}
and Harrington, David M.
and Jaeggli, Sarah A.
and Rast, Mark P.
and Schad, Thomas A.
and Schmidt, Wolfgang
and Mathioudakis, Mihalis
and Mickey, Donald L.
and Anan, Tetsu
and Beck, Christian
and Marshall, Heather K.
and Jeffers, Paul F.
and Oschmann, Jacobus M.
and Beard, Andrew
and Berst, David C.
and Cowan, Bruce A.
and Craig, Simon C.
and Cross, Eric
and Cummings, Bryan K.
and Donnelly, Colleen
and de Vanssay, Jean-Benoit
and Eigenbrot, Arthur D.
and Ferayorni, Andrew
and Foster, Christopher
and Galapon, Chriselle Ann
and Gedrites, Christopher
and Gonzales, Kerry
and Goodrich, Bret D.
and Gregory, Brian S.
and Guzman, Stephanie S.
and Guzzo, Stephen
and Hegwer, Steve
and Hubbard, Robert P.
and Hubbard, John R.
and Johansson, Erik M.
and Johnson, Luke C.
and Liang, Chen
and Liang, Mary
and McQuillen, Isaac
and Mayer, Christopher
and Newman, Karl
and Onodera, Brialyn
and Phelps, LeEllen
and Puentes, Myles M.
and Richards, Christopher
and Rimmele, Lukas M.
and Sekulic, Predrag
and Shimko, Stephan R.
and Simison, Brett E.
and Smith, Brett
and Starman, Erik
and Sueoka, Stacey R.
and Summers, Richard T.
and Szabo, Aimee
and Szabo, Louis
and Wampler, Stephen B.
and Williams, Timothy R.
and White, Charles},
title={The Daniel K. Inouye Solar Telescope -- Observatory Overview},
journal={Solar Physics},
year={2020},
month={Dec},
day={04},
volume={295},
number={12},
pages={172},
issn={1573-093X},
doi={10.1007/s11207-020-01736-7},
url={https://doi.org/10.1007/s11207-020-01736-7}
}

@article{Centeno_2009,
doi = {10.1088/0004-637X/692/2/1211},
url = {https://dx.doi.org/10.1088/0004-637X/692/2/1211},
year = {2009},
month = {feb},
publisher = {The American Astronomical Society},
volume = {692},
number = {2},
pages = {1211},
author = {Centeno, R. and Collados, M. and Bueno, J. Trujillo},
title = {WAVE PROPAGATION AND SHOCK FORMATION IN DIFFERENT MAGNETIC STRUCTURES},
journal = {The Astrophysical Journal}
}

@article{Shchukina_2017,
  author  = {Shchukina, N. G. and Sukhorukov, A. V. and Trujillo Bueno, J.},
  title   = {A Si I atomic model for NLTE spectropolarimetric diagnostics of the 10\,827 \AA\ line},
  journal = {A\&A},
  year    = {2017},
  volume  = {603},
  pages   = {A98},
  doi     = {10.1051/0004-6361/201630236},
  url     = {https://doi.org/10.1051/0004-6361/201630236}
}

@article{Utz_2013,
  author  = {Utz, D. and Jurčák, J. and Hanslmeier, A. and Muller, R. and Veronig, A. and Kühner, O.},
  title   = {Magnetic field strength distribution of magnetic bright points inferred from filtergrams and spectro-polarimetric data},
  journal = {A\&A},
  year    = {2013},
  volume  = {554},
  pages   = {A65},
  doi     = {10.1051/0004-6361/201116894},
  url     = {https://doi.org/10.1051/0004-6361/201116894}
}

@article{Jess_2012,
doi = {10.1088/0004-637X/746/2/183},
url = {https://dx.doi.org/10.1088/0004-637X/746/2/183},
year = {2012},
month = {feb},
publisher = {The American Astronomical Society},
volume = {746},
number = {2},
pages = {183},
author = {Jess, D. B. and Shelyag, S. and Mathioudakis, M. and Keys, P. H. and Christian, D. J. and Keenan, F. P.},
title = {PROPAGATING WAVE PHENOMENA DETECTED IN OBSERVATIONS AND SIMULATIONS OF THE LOWER SOLAR ATMOSPHERE},
journal = {The Astrophysical Journal}
}

@article{Keller_1992,
author={Keller, C. U.},
title={Resolution of magnetic flux tubes on the Sun},
journal={Nature},
year={1992},
month={Sep},
day={01},
volume={359},
number={6393},
pages={307-308},
issn={1476-4687},
doi={10.1038/359307a0},
url={https://doi.org/10.1038/359307a0}
}

@article{Utz_2009,
  author  = {Utz, D. and Hanslmeier, A. and Möstl, C. and Muller, R. and Veronig, A. and Muthsam, H.},
  title   = {The size distribution of magnetic bright points derived from Hinode/SOT observations},
  journal = {A\&A},
  year    = {2009},
  volume  = {498},
  number  = {1},
  pages   = {289--293},
  doi     = {10.1051/0004-6361/200810867},
  url     = {https://doi.org/10.1051/0004-6361/200810867}
}

@article{Yadav_2020,
   title={Vortex flow properties in simulations of solar plage region: Evidence for their role in chromospheric heating},
   volume={645},
   ISSN={1432-0746},
   url={http://dx.doi.org/10.1051/0004-6361/202038965},
   DOI={10.1051/0004-6361/202038965},
   journal={Astronomy &amp; Astrophysics},
   publisher={EDP Sciences},
   author={Yadav, N. and Cameron, R. H. and Solanki, S. K.},
   year={2020},
   month=dec, pages={A3} }

@article{Rempel_2016,
   title={EXTENSION OF THE MURAM RADIATIVE MHD CODE FOR CORONAL SIMULATIONS},
   volume={834},
   ISSN={1538-4357},
   url={http://dx.doi.org/10.3847/1538-4357/834/1/10},
   DOI={10.3847/1538-4357/834/1/10},
   number={1},
   journal={The Astrophysical Journal},
   publisher={American Astronomical Society},
   author={Rempel, M.},
   year={2016},
   month=dec, pages={10} }

@article{Przybylski_2022,
   title={Chromospheric extension of the MURaM code},
   volume={664},
   ISSN={1432-0746},
   url={http://dx.doi.org/10.1051/0004-6361/202141230},
   DOI={10.1051/0004-6361/202141230},
   journal={Astronomy &amp; Astrophysics},
   publisher={EDP Sciences},
   author={Przybylski, D. and Cameron, R. and Solanki, S. K. and Rempel, M. and Leenaarts, J. and Anusha, L. S. and Witzke, V. and Shapiro, A. I.},
   year={2022},
   month=aug, pages={A91} }

@article{Lites_2008,
doi = {10.1086/522922},
url = {https://dx.doi.org/10.1086/522922},
year = {2008},
month = {jan},
publisher = {},
volume = {672},
number = {2},
pages = {1237},
author = {Lites, B. W. and Kubo, M. and Socas-Navarro, H. and Berger, T. and Frank, Z. and Shine, R. and Tarbell, T. and Title, A. and Ichimoto, K. and Katsukawa, Y. and Tsuneta, S. and Suematsu, Y. and Shimizu, T. and Nagata, S.},
title = {The Horizontal Magnetic Flux of the Quiet-Sun Internetwork as Observed with the Hinode Spectro-Polarimeter},
journal = {The Astrophysical Journal}
}

@article{Danilovic_2016,
   title={Internetwork magnetic field as revealed by two-dimensional inversions},
   volume={593},
   ISSN={1432-0746},
   url={http://dx.doi.org/10.1051/0004-6361/201527842},
   DOI={10.1051/0004-6361/201527842},
   journal={Astronomy &amp; Astrophysics},
   publisher={EDP Sciences},
   author={Danilovic, S. and van Noort, M. and Rempel, M.},
   year={2016},
   month=sep, pages={A93} }

@article{Buehler_2019,
  author  = {Buehler, D. and Lagg, A. and van Noort, M. and Solanki, S. K.},
  title   = {A comparison between solar plage and network properties},
  journal = {A\&A},
  year    = {2019},
  volume  = {630},
  pages   = {A86},
  doi     = {10.1051/0004-6361/201833585},
  url     = {https://doi.org/10.1051/0004-6361/201833585}
}

@article{Kahil_2019,
  author  = {Kahil, F. and Riethmüller, T. L. and Solanki, S. K.},
  title   = {Intensity contrast of solar plage as a function of magnetic flux at high spatial resolution},
  journal = {A\&A},
  year    = {2019},
  volume  = {621},
  pages   = {A78},
  doi     = {10.1051/0004-6361/201833722},
  url     = {https://doi.org/10.1051/0004-6361/201833722}
}

@misc{Liu_2025,
      title={Fine-scale opposite-polarity magnetic fields in a solar plage revealed by integral field spectropolarimetry}, 
      author={G. Liu and I. Milić and J. S. Castellanos Duran and J. M. Borrero and M. van Noort and C. Kuckein},
      year={2025},
      eprint={2505.07561},
      archivePrefix={arXiv},
      primaryClass={astro-ph.SR},
      url={https://arxiv.org/abs/2505.07561}, 
}

@misc{BASS2000,
  author       = {{BASS2000 team}},
  title        = {{BASS2000}: a solar spectrum and data service},
  howpublished = {\url{https://bass2000.obspm.fr/solar_spect.php}},
  year         = 2025,
  note         = {Accessed: 2025‐08‐22},
  doi          = {10.25935/yvm9-gk52}
}

@article{Berger_Title_nmbp_2001,
doi = {10.1086/320663},
url = {https://dx.doi.org/10.1086/320663},
year = {2001},
month = {may},
publisher = {},
volume = {553},
number = {1},
pages = {449},
author = {Berger, T. E. and Title, A. M.},
title = {On the Relation of G-Band Bright Points to the Photospheric
Magnetic Field},
journal = {The Astrophysical Journal}
}

@article{ Langhans_nmbp_2002,
	author = {Langhans, K. and Schmidt, W. and Tritschler, A.},
	title = {2D-spectroscopic observations of $\vec G$-band bright structures  
in the solar photosphere},
	DOI= "10.1051/0004-6361:20021196",
	url= "https://doi.org/10.1051/0004-6361:20021196",
	journal = {A&A},
	year = 2002,
	volume = 394,
	number = 3,
	pages = "1069-1076",
}

@article{kannan_2024,
    author = {Kannan, Arjun and Yadav, Nitin},
    title = {Vortex dynamics in various solar magnetic field configurations},
    journal = {Monthly Notices of the Royal Astronomical Society},
    volume = {533},
    number = {3},
    pages = {3611-3622},
    year = {2024},
    month = {08},
    issn = {0035-8711},
    doi = {10.1093/mnras/stae1990},
    url = {https://doi.org/10.1093/mnras/stae1990},
    eprint = {https://academic.oup.com/mnras/article-pdf/533/3/3611/59021696/stae1990.pdf},
}








\end{document}